\documentclass[12pt, reqno, oneside,amsart]{article}
\usepackage{geometry}
\usepackage{hyperref}
\hypersetup{
	colorlinks = true,
	linkcolor=cyan,
	citecolor=red,
	urlcolor=magenta,
	menucolor=magenta
	}

\usepackage{cite}

\usepackage{amsmath}
\usepackage[normalem]{ulem}
\input epsf

\usepackage[shortlabels]{enumitem}

\usepackage{graphicx}
\usepackage{amsfonts}
\usepackage{amssymb}

\usepackage{etoolbox}
\usepackage{titlesec}
\usepackage{titletoc}
\usepackage{titletoc}

 \usepackage[utf8]{inputenc}

\usepackage[titletoc]{appendix}

\usepackage{enumitem}

\usepackage{empheq}

\usepackage{mathrsfs}  

\usepackage{dsfont}

\usepackage{stmaryrd}

\usepackage{subfigure}

\def\m{\mu}
\def\n{\nu}
\def\a{\alpha}

\def\s{\sigma}

\def\la{\lambda}
\def\La{\Lambda}

\def\ka{\varkappa}

\def\ga{\gamma}

\def\rm{\mathrm}
\def\cal{\mathcal}
\def\scr{\mathscr}

\def\pa{\partial}

\def\tQFT{\widetilde{\rm{QFT}}}

\def\QFT{\rm{QFT}}

\def\DW{\rm{DW}}

\def\tDW{\widetilde{\rm{DW}}}

\def\c{c}

\def\UV{\rm{UV}}

\def\IR{\rm{IR}}

\def\onsh{\text{on-sh}}

\def\be{\begin{equation}}
\def\ee{\end{equation}}
\def\br{\begin{eqnarray}}
\def\er{\end{eqnarray}}
\def\bsub{\begin{subequations}}
\def\esub{\end{subequations}}

\titleformat{\section} 
[hang]                 
	{
	\filcenter
	\normalsize
	\large
	} 
{\thesection.}       
{0.3em}                  
{}

\titlespacing{\section}
{0pt}                  
{20pt}                  
{13pt}                  

\titleformat{\subsection} 
[hang]                 
{
\itshape
\filcenter
} 
{\thesubsection.}       
{0.3em}                  
{}                     

\titlespacing{\subsection}
{0pt}                  
{20pt}                  
{15pt}                  

\titleformat{\subsubsection} 
[hang]
{
\filcenter
\normalsize
\itshape
} 
{\thesubsubsection.}    
{0.3em}                
{}                     

\titlespacing{\subsubsection}
{0pt}                  
{25pt}		        
{15pt}                  

\usepackage{authblk}

\title{Conformal Scale Factor Inversion for Domain Walls and Holography}

\author{A.A. Lima\thanks{andrealves.fis@gmail.com}}
\author{U. Camara da Silva\thanks{ulyssescamara@gmail.com}}
\author{G.M. Sotkov\thanks{gsotkov@gmail.com}}

\affil{\small \textit{Department of Physics, Federal University of Esp\'irito Santo, 29075-900 \\ Vit\'oria, Brazil}}

\begin{document}

\pagenumbering{gobble}

\maketitle

\begin{abstract}

 We describe a correspondence between domain wall solutions of Einstein gravity with a single scalar field and self-interaction potential. The correspondence we call `conformal scale factor inversion (CSFI)' is a map comprising the inversion of the scale factor in conformal coordinates, and a transformation of the field and its potential which preserves the form of the Einstein equations for  static and isotropic domain walls.  By construction, CSFI maps the asymptotic AdS boundary to the vicinity of a naked singularity in a theory with a special Liouville (exponential) scalar field potential; it is also a map in the parameter space of exponential potentials.
The correspondence can be extended to linear fluctuations, being akin to an S-duality, and can be interpreted in terms of `SUSY quantum mechanics' for the fluctuation modes.
The holographic implementation of CSFI relates the UV and IR regimes of a pair of holographic renormalization group flows; in particular, it is a symmetry of the GPPZ flow.

{\footnotesize 
\bigskip
\noindent
\textbf{Keywords:} 

\noindent
Domain wall solutions, gauge/gravity correspondence, holographic RG flow.
}

\end{abstract}

\maketitle

\newpage

\tableofcontents

\newpage

\pagenumbering{arabic}

\section{Introduction}

Domain wall solutions of (super)gravity are important both from the gravitational point of view \cite{Cvetic:1996vr} and because they correspond to the holographic renormalization group flows of dual QFTs \cite{Akhmedov:1998vf,Skenderis:1999mm,Boonstra:1998mp,deBoer:1999tgo,DeWolfe:1999cp,ArkaniHamed:2000ds,Bianchi:2001de,Bianchi:2001kw,Papadimitriou:2004rz,Papadimitriou:2004ap}.
A domain wall solution with $d$-dimensional Poincar\'e symmetry has the form
\be
d s^2 = a^2(z) \left[ dz^2 + \eta_{ab} dx^a dx^b \right] . 
\label{ConfCoordDWa}
\ee
Large values of the scale factor $a(z) \to \infty$ correspond to the UV limit of the holographic theory, small values of $a \to 0$ to the IR, and the profile of the function $a(z)$ describe different ways in which the holographic RG can flow from the UV to IR. 
In the standard case \cite{Papadimitriou:2004ap,Bianchi:2001kw,Bianchi:2001de,deBoer:1999tgo} the UV is a fixed point corresponding to the boundary of an asymptotically $AdS_{d+1}$ geometry, and conformal symmetry of the $\rm{QFT}_d$ is broken by a (self-interacting) scalar field $\phi$, that drives the RG flow to the IR. The simplest fate of the flow is a IR fixed point with a different $AdS_{d+1}$ asymptotics near the AdS horizon, but there are other explicit constructions of flows with non-trivial and interesting IR behavior --- confinement, screening, etc. --- such as the GPPZ example of deformations of $\cal N = 4$ SYM \cite{Girardello:1998pd,Girardello:1999hj,Girardello:1999bd}. Typically, these involve singularities \cite{Gubser:2000nd}; in general, a bottom-up approach \cite{Gursoy:2007cb,Gursoy:2007er,Kiritsis:2016kog,Megias:2014iwa,Megias:2015nya} has proven to be fruitful for the classification of all possible different behaviors of  solutions.

In the present paper, we introduce a map between the UV and IR limits of pairs of domain wall solutions related by inversion of the conformal scale factor $a(z)$. More precisely, the map, which we call \emph{`conformal scale factor inversion (CSFI)'}, relates two domain wall geometries, $\DW$ and $\tDW$, in Einstein gravity, with scalar fields $\phi$ and $\tilde \phi$ and potentials $V(\phi)$ and $\tilde V(\tilde \phi)$, 
\be
 a(z) \mapsto \tilde a(\tilde z) = \frac{\c^2}{a(z_0 \pm z)} ,
\quad
\phi \mapsto \tilde  \phi(\phi) ,
\quad
V(\phi) \mapsto \tilde V (\tilde \phi) ,
	\label{SFDintroMesmo}
\ee
$\c$ is a free parameter. 
The transformation is such as to preserve the form of the Einstein equations, ensuring that if $a(z), \phi(z)$ solve the equations with the potential $V(\phi)$, then $\tilde a(\tilde z), \tilde \phi(\tilde z)$ solve the equations with $\tilde V(\tilde \phi)$.

The inversion of the scale factor \emph{in conformal coordinates} can be seen as a kind of discrete $Z_2$ Weyl transformation.
The transformation of the scalar field and its potential is in fact rather non-trivial.
It maps (asymptotically) AdS solutions to the (asymptotic) solutions of a special Liouville exponential potential $V = V_0 e^{v \ka \phi}$ with $v = 2(d-1)^{-1/2}$. It also maps different Liouville models by changing the parameters $v \mapsto \tilde v$ in a definite manner.
Domain walls with exponential potentials are one of the few cases, outside of asymptotically AdS geometries, for which holography has a precise quantitative formulation \cite{Kanitscheider:2008kd}: they can be described as AdS-linear dilaton solutions dual to QFTs which are not conformal, but possess a `generalized conformal structure'. They can also be obtained from dimensional reductions of pure AdS solutions in a larger number of dimensions \cite{Kanitscheider:2009as,Gouteraux:2011qh}.

In principle, given any (as long as some restrictions are observed) solution $\DW$ corresponding to the RG flow of a $\QFT$, the CSFI map will yield an image solution $\tDW$ and the flow of a different $\tQFT$.
The most important holographic information contained in homogenous, isotropic backgrounds is the beta function  $\beta(\phi) = - d\phi / d \log a$ describing the RG flow of the coupling $\phi$ with the energy scale $\log a$. The  explicit CSFI transformations of $\beta$ and of the remaining RG data turn out to have quite a simple form. 
Non-trivial IR limits due to exponential behavior of $V(\phi)$ are important in phenomenological applications 
\cite{Gursoy:2007cb,Gursoy:2007er,Gursoy:2016ggq,Gubser:2008yx,Gouteraux:2011ce, Gouteraux:2011qh}, and a natural question is whether these IR limits are related by CSFI to some interesting UV limit. 
As we will show, this is indeed the case. In particular, we can construct a class of models which are \emph{invariant} under (\ref{SFDintroMesmo}), and one of these models is the well-known GPPZ flow \cite{Girardello:1999bd} (truncated to a single scalar field). 

Beyond isotropic backgrounds, CSFI can be consistently extended to fluctuations, like the S-duality of cosmological fluctuations \cite{Brustein:1998kq}.
The equations for tensor modes around any domain wall present a symmetry under CSFI relating the modes and their conjugate momenta; the equations for the scalar modes posses the same symmetry in theories with exponential potentials (but only in those theories). The symmetry can be interpreted in terms of the well-known ``SUSY quantum-mechanical'' \cite{Cooper:1994eh} description of domain wall fluctuations.
A given pair of modes related by CSFI are superpartners, each can be found from acting with the SUSY generators on the other. This relates the corresponding ``wave functions'' in such a way that fixing the boundary conditions of fluctuations in one model fixes them in its CSFI-pair as well, even in cases where the extension of CSFI is only asymptotic, such as in asymptotically Liouville/AdS backgrounds.

The structure of the paper is as follows.
In Sect.\ref{SectDWand} we review the description of isotropic domain walls in terms of a system of first-order differential equations.
In Sect.\ref{SectSFD} we introduce CSFI, describe its properties and construct explicit examples of pairs of solutions, focusing in asymptotically AdS/Liouville geometries. 
In Sect.\ref{SectSFDasdulflows}, we discuss pairs of solutions from the point of view of holographic RG flows, and in Sect.\ref{SectRenorActio} we discuss the relations between one-point functions. 
In Sect.\ref{SectFluctuandSpec} we describe the extension of CSFI to fluctuations and the consequences for wave-functions and spectra of $d$-dimensional eigenstates. 
In Sect.\ref{SectConclusion} we conclude with a discussion of the CSFI correspondence, and some of its possible developments. 
Secondary topics and examples are left for the appendices.

\section{Domain walls}	\label{SectDWand}

Let us briefly review the basic features of domain walls in Einstein gravity coupled to one scalar field  \cite{Cvetic:1996vr, Cvetic:1994ya}. 
We work in the Einstein-frame, with action
\be
S =  \int d^{d+1} x \ \sqrt{-g} \Bigg[ \frac{1}{\ka^2} R - \frac{1}{2} g^{\m\n} \pa_\m \phi \pa_\n \phi - V(\phi) \Bigg] ,
\label{ActSGVph}
\ee
and consider static and flat domain walls with $d$-dimensional Poincar\'e symmetry%
\footnote{%
$\eta_{ab}$ is the Minkowski metric, where $a,b = 0,1,2, \cdots, d-1$ denote $d$-dimensional ``transversal'' coordinates and $\mu,\nu$ bulk coordinates. Newton's constant in $(d+1)$ dimensions is (proportional to) $\ka^2$.}
\begin{align}
\begin{split}
d s^2 &= e^{2A(z)} \left[ dz^2 + \eta_{ab} dx^a dx^b \right] , \quad e^{A(z)} \equiv a(z) ,
\\
 \phi &=  \phi (z) . 
\end{split}
\label{ConfCoordDW}
\end{align}
The field equations obtained from the action (\ref{ActSGVph}) have the form
\bsub
\begin{align}
2 A'' + (d-2) A'^2 = - \frac{\ka^2}{d-1}  \left[ \tfrac{1}{2}  \phi'^2 + e^{2A} V( \phi) \right]	\label{AppEq}
\\
 \phi'' + (d-1) A'  \phi' = e^{2A} \pa_{ \phi} V(  \phi ) \label{KGordz}
 \\
 d (d-1) A'^2 - \tfrac{1}{2} \ka^2  \phi'^2 = - \ka^2 e^{2A} V (  \phi) \label{HamilConstz}
\end{align} \label{EinsKGzfds}\esub
where $' \equiv d / dz$. 
Eqs.(\ref{EinsKGzfds})  can be rewritten in an equivalent form as the first-order system 
\bsub
\begin{align}
A'(z)  &= - \frac{\ka}{d-1} e^{A(z)} W(\phi)
			\label{EinsEqHom2Z}
\\
 \phi'(z)  &= \frac{2}{\ka} e^{A(z)} \frac{d W (\phi)}{d\phi}  
\label{EinsEqHom3Z}
\end{align}
\label{EinsEqHombothZ}\esub
by introducing  an (auxiliary) function $W(\phi)$ called the `superpotential' due to its  (fake or true) supersymmetric origin
\cite{deBoer:1999tgo,Skenderis:1999mm,Townsend:1984iu,Freedman:2003ax}.
The constraint (\ref{HamilConstz}) determines  the potential $V(\phi)$ in terms of the superpotential:
\be
V( \phi ) = \frac{2}{\ka^2} \left[ \frac{d W (\phi)}{d\phi } \right]^2 - \frac{d}{d-1} W^2 (\phi),
					\label{V1M}
\ee
or vice-versa --- when $V(\phi)$ is given, one should solve Eq.(\ref{V1M}) for $W(\phi)$ and then to use this (non-unique)  solution for solving the first-order system. We are  mostly interested  in the proper supergravity DWs of BPS-type, where the superpotential is  an important input  and the first-order system is nothing but the integrability consistency conditions for the BPS Killing spinor equations. 
Notice that in the patches of spacetime where   $W(\phi)$ is monotonic, one can use $\phi$ itself as a coordinate and find $A = A(\phi)$ from the equation $d A / d\phi = - 1/\beta(\phi)$, where
\be
\beta (\phi ) =  - \frac{d \phi}{d A} =  \frac{2 (d-1)}{\ka^2} \ \frac{\pa_\phi W (\phi) }{ W(\phi) } ,
			\label{betaM}
\ee
which has the holographic interpretation of the beta-function for the running coupling $\phi$  in the dual QFT (cf. Sect.\ref{SectHoloImpl}).

\section{Conformal scale factor inversion correspondence} 	\label{SectSFD}

We want to construct a map between two domain wall solutions 
--- which we call $\DW$ and $\tDW$ --- 
such that their scale factors are related by an inversion,
\bsub
\be
\tilde a(\tilde z) = \frac{\c^2}{a(z)} \, , \qquad \tilde z = \pm z + \rm{constant} 	\label{sfacinvers}
\ee
where $c$ is a dimensionless parameter.
The motivation for finding such a map is simple: it naturally relates a near-singular  geometry ($\tilde a /c \ll 1$) to a small-curvature geometry ($a/c \gg 1$). 
(The causal structure of spacetimes related by (\ref{sfacinvers}) is described in App.\ref{AppCausalStruc}.)
Holographically, the correspondence relates the deep IR of the RG flow to the UV limit of a (as we will see) different RG.
Since we want to relate two domain wall solutions, the inversion of the scale factor (\ref{sfacinvers}) must preserve the form of the Einstein equations (\ref{EinsEqHombothZ}). 
It is straightforward to see that this is only true if the scalar \emph{and the potential} transform in a specific way, as
\be
\begin{split}
\tilde \phi'^2 &= - \phi'^2 + \frac{\ka^2}{d-1} a^2(\phi) W^2(\phi) \, ,
\\
\tilde V(\tilde \phi) &= - \frac{a^4(\phi)}{\c^4}  \left[ V(\phi) + 2 W^2(\phi) \right] .
\end{split}
\label{SFDphiTrans}
\ee\label{SFinvfo}\esub
Hence we are relating by scale-factor inversion a pair of solutions in \emph{two different theories}, differing by the scalar potential.  

We are going to call the map composed by the three transformations (\ref{SFinvfo}) a \emph{`conformal scale factor inversion'} (CSFI) correspondence.

We can also express the map in terms of the superpotentials $W(\phi)$ and $\tilde W(\tilde \phi)$ of the related theories.
Scale-factor inversion implies that the superpotential transforms as
\be
 \tilde a (\tilde \phi) \tilde W (\tilde \phi) = \mp a(\phi) W(\phi) 
 \quad \text{or} \quad  \tilde W (\tilde \phi) =  \mp \frac{a^2(\phi)}{\c^2} W(\phi), \label{WatsnfSFD} 
\ee
with the signs following those of (\ref{sfacinvers}). We can further  set $\c = 1$ by choosing the normalization of the scale factors in the two solutions.

The field transformation $\phi \mapsto \phi(\tilde \phi)$ can be written in a more direct form by using the scale factor $a(\phi)$, as a function of the scalar field,  together with Eq.(\ref{betaM}), 
\be
\tilde \phi (\phi ) = \tilde \phi_0 \pm \int_{\phi}^{\phi_0} d \phi \Bigg[ \frac{4(d-1)}{ \ka^2 \beta^2 (\phi)} -1 \Bigg]^{1/2} .
			\label{sigma_dual2M}
\ee
The condition that r.h.s. has to be  real introduces the upper bound 
\be
\beta^2 (\phi) \leq 4(d-1)/\ka^2 .
		\label{dualbetabound}
\ee 
If this bound is violated in a given solution $\DW$, the kinetic energy of $\tilde \phi$ has the wrong sign, and the solution $\tDW$ is unstable.
In other words, the bound (\ref{dualbetabound}) preserves the null energy condition (NEC), which for the single homogeneous scalar field reads $\phi'^2(z) \geq 0$. It is not difficult to  derive the transformation of the beta-functions,
\be
\tilde \beta (\tilde \phi) = \pm \Bigg[ \frac{4(d-1)}{ \ka^2 \beta^2 (\phi)} -1 \Bigg]^{1/2}  \beta(\phi)
	\label{dualbetapm}
\ee
where the signs follow the ones in (\ref{sigma_dual2M}).
This can be written in a symmetric way (that however hides the relation between signs) as
\be
\beta^2 (\phi) + \tilde \beta^2 (\tilde \phi ) = 4(d-1)/ \ka^2 ,
						\label{dualM}
\ee
and from the symmetry of this equation one sees that if (\ref{dualbetabound}) holds for $\DW$ it holds for $\tDW$ as well.

The ambiguities in sign appearing in Eqs.(\ref{WatsnfSFD}),  (\ref{sigma_dual2M}) and (\ref{dualbetapm})
 correspond to different ambiguities in Eq.(\ref{V1M}), which determines $W(\phi)$ from $V(\phi)$. To be definite, in what follows we adopt the following conventions: 
\begin{enumerate}[i)]
	
	\item	\label{signsoW}
	\emph{We always choose $\tilde z = - z + \rm{const.}$ in (\ref{sfacinvers}). 
	Thus both $\tilde W > 0$ and $W > 0$, i.e. the \textbf{plus} sign is chosen in (\ref{WatsnfSFD}).} Choosing a definite sign of the superpotential has no physical consequence, it only stipulates that both $a$ and $\tilde a$ grow with their respective radial coordinates. 
	
	\item \label{itemiisidW}
	\emph{We always choose the \textbf{plus} sign in Eqs.(\ref{sigma_dual2M}) and (\ref{dualbetapm}).}
	This particular sign ambiguity stems from invariance of Eq.(\ref{V1M}) under a change of sign of $d\tilde W / d\tilde\phi$. Once we have fixed $\tilde W > 0$ in item \ref{signsoW}, the sign of $\tilde \beta$ is fixed by $d\tilde W / d\tilde\phi$, cf. Eq.(\ref{betaM}).
	A choice of the opposite sign would select a different $\tilde W(\tilde \phi)$, but in a theory with the same  $\tilde V(\tilde \phi)$.%
	\footnote{See \cite{Kiritsis:2016kog} for an extensive discussion of the solutions of the superpotential equation (\ref{V1M}).}
		\end{enumerate}

Some observations are in order.	

There is a well-known relation between domain walls and FRW spacetimes. 
In cosmology, the maps obtained by inversion of the scale factor are known as `scale factor dualities', in analogy to Veneziano's work  \cite{Veneziano:1991ek}.  CSFI corresponds to the inversion of the FRW scale-factor in conformal time developed by the present authors in \cite{dS:2015fwa,dS:2016kdc,Sotkov:2016mks}.

In this paper we use only the first-order system (\ref{EinsEqHombothZ}), but its easily verified that the transformations (\ref{SFinvfo}) actually preserve \emph{the full second-order equations (\ref{EinsKGzfds}).}
(See \cite{dS:2015fwa}.)
Starting from the second-order equations, it is easy to see that preservation of the first-order system automatically follows.
The first-order system, i.e. the superpotential, exists whenever a solution $a(z)$ is a (piecewise) monotonic function of $z$, and the converse is also  true \cite{DeWolfe:1999cp}. 
Now, if $a(z)$ is monotonic, then $1/a(z)$ is monotonic as well. Hence CSFI maps two monotonic functions which must be described by the corresponding superpotentials.
	
The map, however, is not a symmetry of the action --- only of the field equations.
Dismissing a volume factor $\int d^dx$, the action for the ansatz (\ref{ConfCoordDW}) can be put in the well-known (pseudo)-BPS form \cite{Skenderis:1999mm,DeWolfe:1999cp}
\be
\begin{split}
S &=    \int_{z_1}^{z_2} \! dz  \ a^{d} \Bigg[  \frac{d (d-1)}{\ka^2} \left( \frac{a'}{a}  + \frac{\ka a W}{d-1}  \right)^2	
	- \frac{1}{\ka}  \left(\frac{\ka}{2a} \phi'  -  a  \frac{ d W }{d\phi }  \right)^2  \Bigg]
\\
	&\quad -  \Bigg[\frac{2 d}{\ka^2}  \, a^{d-1} \left( \frac{a'}{a} +  \frac{\ka}{d} a W  \right) \Bigg]_{z_1}^{z_2}	
\end{split}					\label{ActforBPSnoonConf}
\ee
which after the CSFI transformations (with $dz=\pm d\tilde z$ and $\c = 1$) become
\be
\begin{split}
S &= \pm  \int_{\tilde z_1}^{\tilde z_2} \!\! d\tilde z  \ \frac{1}{ \tilde a^d } \Bigg[ \frac{d (d-1)}{\ka^2} \left( \frac{\tilde a'}{\tilde a} + \frac{\ka \tilde a \tilde W}{d-1}  \right)^2 
	- \frac{1}{\ka} \left( \frac{d \tilde \phi}{d \phi} \right)^2  \left(\frac{\ka}{2\tilde a} \tilde \phi'  - \tilde a  \frac{ d \tilde W }{d \tilde \phi }  \right)^2  \Bigg]
\\	
	&\quad - \Bigg[ \frac{2 d}{\ka^2}  \, \frac{1}{ \tilde a^{d-1}} \left( \frac{\tilde a'}{\tilde a} +  \frac{\ka}{d} \tilde a \tilde W  \right) \Bigg]_{\tilde z_2}^{\tilde z_1}  		
\end{split}
\ee
The BPS-like separation of squares is preserved, but the position of $a$ and $\tilde a$ is different on the overall coefficients. Hence there is no invariance, not even on-shell, when only the boundary terms survive.

We are considering here only flat domain walls (\ref{ConfCoordDW}) but solutions with $\eta_{ab}$ in (\ref{ConfCoordDW}) replaced by a metric $g_{ab}$ with constant curvature proportional to the cosmological constant $\La_d$ are also important \cite{DeWolfe:1999cp}.
A remarkable feature of CSFI is that the second-order equations (\ref{EinsKGzfds}), modified by the presence of $\La_d$ terms, are also invariant under the \emph{same} transformations (\ref{SFinvfo}).
That is, \emph{CSFI is valid for any domain wall with maximally symmetric slices.}
This can be seen by looking at the first-order system for the curved domain-walls. As shown in \cite{DeWolfe:1999cp}, Eqs.(\ref{EinsEqHombothZ}) are modified by the appearance of a factor $\ga \equiv \left[ 1 - \a \La_d / (a W)^2 \right]^{1/2}$, where $\a$ is a numerical factor \cite{DeWolfe:1999cp}. This factor is invariant under CSFI, because of (\ref{WatsnfSFD}).

In our definition of the CSFI map, the emphasis in `conformal'  comes from $a(z)$ being a conformal factor. 
This point is important because implementing a scale factor inversion in a different coordinate leads to a different map, with a different field transformation. For example, suppose we implement the scale factor inversion $\tilde a(r) = \c / a(r)$, where $r = \int dz \ a(z)$. 
This gives a \emph{different} map, with field transformations such that $\dot \phi^2(r) \mapsto - \dot \phi^2(r)$.%
\footnote{This is easy to see from the Einstein equations, since now we are imposing $A(r) \mapsto - A(r)$.}
 Thus in this alternative correspondence the kinetic energy of the scalar field always changes sign and the NEC is always broken. The equivalent duality in Friedmann-Robertson-Walker spacetimes \cite{Dabrowski:2003jm} can be used to relate standard and phantom cosmological models \cite{Dabrowski:2003jm, Chimento:2003qy}.

One can concoct other maps, by using other ``lapse functions'' $N(z)$ on the domain wall metric (\ref{ConfCoordDW}), although the most natural ones are either CSFI, with $N(z) = a(z)$, or the map just mentioned with $N = 1$.
We have seen that the latter always violates the null energy condition by construction, while CSFI allows for pairs $\DW$ and $\tDW$ both obeying the NEC.  Also, CSFI is the only such map, for whatever choice of $N(z)$, which can be extended for curved maximally symmetric solutions. 
But there are two more remarkable features of CSFI. First, it relates asymptotically AdS potentials to exponential potentials, and is also a map on the parameter space of exponential-potential models where it acts as a strong-/weak-coupling map. Second, CSFI can be naturally extended to fluctuations in these cases. Remarkably, AdS and exponential-potential models are two classes of theories with a precise holographic interpretation. The remaining of this paper is devoted to explore these features.

\section{Solutions related by CSFI}

We now proceed to discuss explicit realizations of the CSFI transformations (\ref{SFinvfo}). Of course, the obvious place to start would be with AdS solutions: what is their image under CSFI? The answer is a domain wall in a special Liouville model, with a specific exponential potential. 
In fact, it turns out that the CSFI transformations are realized in a particularly simple way in Liouville models, acting as a map in the parameter space of the exponential potentials, of which the AdS solution can be obtained from a limiting procedure.
We therefore start, in \S\ref{SectLiouvandAdS}, by describing these maps. 

Exponential potentials can be obtained in different ways, from an AdS linear dilaton solution of a Jordan-frame action, and from consistent dimensional reductions of a higher-dimension pure AdS solution. In \S\S\ref{SectNonConfpBrn}-\ref{SectNonConfpBrn2} we analyze how CSFI can be interpreted in these frameworks.

In \S\ref{SectAsumptAdSanddual}, we give as a further example the more complicated case of the image of a standard asymptotically AdS potential. Finally, we give an example of a non-trivial theory (and a solution) invariant under the transformations.

\subsection{Liouville potentials and AdS space} \label{SectLiouvandAdS}

Consider Liouville, i.e. exponential, potentials 
\begin{align}
&W(\phi) = \frac{2/v^2}{\ka L} e^{v \ka \phi / 2 } 
		\label{W0expv}
\\
\begin{split}
&V(\phi) = -  \frac{(d-1) (v_c/v)^4}{\ka^2 L^2} \Big[  d - (v/v_c)^2   \Big] e^{v \ka \phi } \ , 
\\
&
\quad  v_c \equiv \sqrt{\frac{2}{d-1}} .
		\label{VdephiExpW}	
\end{split}			
\end{align}
Here $v > 0$ is a dimensionless parameter. The amplitude has been chosen appropriately to simplify the scale factor in terms of a length scale $L > 0$. 
We assume that $V \leq 0$,  hence
\be
0 < v <  \sqrt d \ v_c .
	\label{vvmaxbonud} 
\ee
The beta-function given by (\ref{betaM}) is a constant parameterized by $v$,
\be
\beta =   (d-1) v / \ka 
		\label{betaexpcWc}
\ee
so integrating Eq.(\ref{betaM}) we find the $\DW$ solution
\be
a(\phi) =  \exp \left[ - \frac{\ka \phi}{ (d-1) v }\right] ,
	\label{aofphexpW}
\ee
while integration of the first-order system (\ref{EinsEqHombothZ}) yields
\begin{align}
&a(z) = \left[ \tfrac{(v^2 - v_c^2 )(z_0 - z)}{v^2 L } \right]^{\frac{v_c^2}{v^2 - v_c^2}} 
&&
		\Bigg\{\!\!\!\!\!\!\!\!
		\begin{split}
		&\text{$v_c < v$,  $z \in (-\infty, z_0)$}
	\\
		&\text{$v_c > v$,  $z \in (z_0, +\infty)$}
		\end{split}
		&&
\label{SolforazExp}
\\
&a(z) =  \exp \Big[  \frac{z_0 - z}{L} \Big]
&&
		&\text{$v_c = v$,   $z \in (-\infty, \infty)$}
		&&		
\label{SolforazExp2}		
\end{align}
where $z_0$ is an integration constant.	
In all cases there is a singularity when $a \to 0$, but with one  crucial difference: for $v > v_c$  the singularity is reached at the \emph{finite} radius $z_0$. For $v \leq v_c$ the point $z_0$ is where the scale factor diverges, and the singularity is at $z = + \infty$.
The solutions for $\phi$ are
\begin{align}
& \phi(z) = - \frac{2}{\ka v} \log \Bigg[ \left[ \frac{(v^2 - v_c^2)(z_0 - z)}{v^2L}  \right]^{\frac{v^2}{v^2-v_c^2}} \Bigg]
&& \text{if $v \neq v_c$}
		\label{phipfzExpsw}
\\		
& \phi(z) = - \frac{2}{\ka v} \frac{(z_0 - z)}{L} 
&&
		\text{if $v = v_c$}	\label{phipfzExpswCr}
\end{align}
with ranges of  $z$ again as in (\ref{SolforazExp})-(\ref{SolforazExp2}). 
Note that the singularity is always at $\phi \to + \infty$; cf. (\ref{aofphexpW}).

The CSFI image $\tDW$ of this solution has $\tilde a = \c^2 / a$. For $v \neq v_c$, 
\begin{align}
\begin{split}
& \tilde a(\tilde z) = \left[ \frac{(v^2 - v_c^2) (\tilde z_0 - \tilde z)}{v^2 \tilde  L } \right]^{- \frac{v_c^2 }{ v^2 - v_c^2}} ,
\\
& \tilde z - \tilde z_0 = - (z - z_0) ,
\end{split}			\label{tildeazqds}
\end{align}
for some constant $\tilde L$. Eq.(\ref{tildeazqds}) has the same structure as (\ref{SolforazExp}), but with
\be
\tilde v^2 - v_c^2 = v_c^2 - v^2	. \label{qtqvtvtranf}
\ee
Since the scale factors have the same form, the potential and the superpotential must again be  exponentials.
The field $\tilde \phi$ is given by Eq.(\ref{sigma_dual2M}), whose integral is trivial, 
\be
\tilde \phi / \tilde v = - \phi / v .	\label{scfFileLiouvSF}
\ee
We have chosen the plus sign in (\ref{sigma_dual2M}), and also set the integration constant to zero.
With this choice, from Eq.(\ref{WatsnfSFD}), we find that the potential and superpotential are
\begin{align}
\begin{split}
 \tilde W( \tilde \phi) &= \tilde W_0 e^{ \tilde v \ka \tilde \phi / 2 } \ ,
\\
 \tilde V(\tilde \phi) &= -  \frac{\tilde W_0^2}{d-1} \left(  d - \frac{\tilde v^2}{v_c^2}   \right) e^{ \tilde v \ka \tilde \phi } .
 \end{split}
\end{align}
Note that we have the map $\{\phi \to -\infty \} \mapsto \{\tilde \phi \to + \infty\}$, corresponding, from Eq.(\ref{aofphexpW}) to $\{ a \to \infty \} \mapsto \{\tilde a \to 0 \}$ (and vice-versa).

\begin{figure}[t] 
\centering
\includegraphics[scale=0.3]{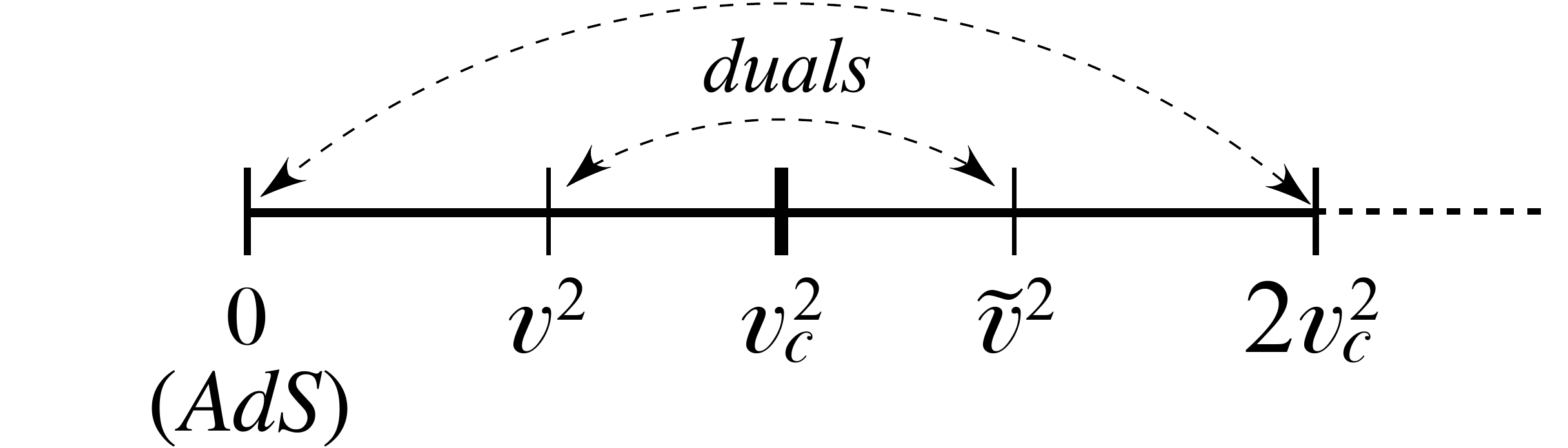}
\caption{Parameter space of pairs of domain walls with exponential potentials related by CSFI.}
\label{RangeOfExp}
\end{figure}

Thus we have found that CSFI is a map between exponential potentials:
\begin{align}
 V(\phi) = V_0 e^{v \ka \phi} \quad &\xleftrightarrow{\quad \rm{CSFI} \quad} \quad \tilde V(\tilde \phi) = \tilde V_0 e^{\tilde v \ka \tilde \phi} 
 	\label{VxpVexpSF}
\\	
0 < v^2 < v_c^2 \quad &\xleftrightarrow{\quad \rm{CSFI} \quad} \quad v_c^2 < \tilde v^2 < 2 v_c^2 
	\label{SFDwsaprangv}
\end{align}
The correspondence reflects a value of $v^2$ across the critical $v_c^2$, as shown in Fig.\ref{RangeOfExp}. Then $v^2 = 0$ is mapped to $\tilde v^2 = 2 v_c^2$, and  any $v^2 > 2 v_c^2$ is mapped to a $\tilde v^2 < 0$. So requiring that the pair $(v, \tilde v)$ is real introduces the bound
\be
0 \leq v^2 \leq 2  v_c^2
\qquad 
\text{or}
\qquad
v^2 \leq 4 / (d-1) .
			\label{bongvvc}
\ee
Eq.(\ref{qtqvtvtranf}) is equivalent to the beta-function transformation (\ref{dualM}) and the bound (\ref{bongvvc}) corresponds to the bound (\ref{dualbetabound}). 
As discussed, a $\DW$ violating the bound has the corresponding $\tDW$ with the wrong sign of the kinetic term.

%
%

The critical value $v = v_c$ gives the simplest example of an `invariant potential', symmetric under CSFI.
A Liouville model with exponential parameter 
\be
v = \sqrt2 v_c = 2 / \sqrt{d-1} ,
\ee
for which (\ref{bongvvc}) is saturated, will be called a \emph{`special Liouville'} solution. The special nomenclature is because the corresponding parameter is $\tilde v = 0$, which gives not an exponential potential anymore but a (negative) constant $\tilde V$, hence \emph{the pair of the special Liouville is AdS space.} 
Note however that, as $v \to 0$ the solutions for $a$ and $\phi$ must be looked at carefully. 
In particular, with our chosen parameterization for the superpotential (\ref{W0expv}), the limit $v \to 0$ must be taken with the fixed product
$v^2 L \to 2\ell / (d-1)$, where $\ell$ is the radius of AdS.
For example, Eq.(\ref{aofphexpW}) does not have a limit, because in pure AdS there is no scalar field; but Eq.(\ref{SolforazExp}) does have the right limit, giving $a = \ell / z$.

\subsubsection{Linear dilaton solutions} 	\label{SectNonConfpBrn}

The Liouville solutions obtained above are related to a linear-dilaton AdS solution of the Jordan-frame action 
\be
\begin{split}
S =  \int \!\! d^{d+1} x \, \sqrt{-G} \ & \frac{e^{\Phi}}{\ka^2} \Bigg[ \scr R 
	+ \Big(1 + \tfrac{1}{2\a} \Big)  G^{\m\n} \pa_\m \Phi \pa_\n \Phi + C \Bigg] 
	\label{EuclStraga}
\end{split}	
\ee
where $\scr R = \scr R(G)$ is the Ricci scalar for the metric $G_{\m\n}$, and $C$ is a constant.
This action can describe the decoupling limit of D$p$-branes for some specific values of $\a$ and $C$ \cite{Kanitscheider:2008kd,Boonstra:1998mp} but we can, and will, consider these parameters to be arbitrary. The Einstein-frame metric $g_{\m\n}$ is found with a conformal transformation 
and by defining the canonically normalized field $\phi$,
\be
\begin{split}
g_{\m\n} &= \exp \Big(  \frac{2 }{d-1} \Phi \Big) G_{\m\n} ,
\\
\ka \phi &= - \sqrt{2} \ \sqrt{\frac{2\a - (d-1)}{2\a(d-1)} } \;  \Phi ,
\end{split}
	\label{StritoEsimetStritoEsidil}
\ee
so that (\ref{EuclStraga}) assumes the standard Einstein-Hilbert form (\ref{ActSGVph}) with a potential $V(\phi) =  - \frac{1}{\ka^2} \, C \exp \left( - \tfrac{2}{d-1}   \Phi (\phi) \right)$ that reads
\be
V(\phi) =  - \frac{1}{\ka^2} \, C \exp \Bigg[   \sqrt{ \frac{4 \a}{(d-1)(2\a + 1-d)}} \ka \phi \Bigg] .
	\label{Vphiagakaph}
\ee
The exponent is uniquely parametrized by $\a$. We can put it in the same form as (\ref{VdephiExpW}) by defining
\be
v =  \sqrt{ \frac{4\a}{(d-1)(2\a + 1-d)} } ,
\quad
v_c = \sqrt{ \frac{2}{d-1} } .
	\label{vsqtrage}
\ee
Solving (\ref{vsqtrage}) for $\a$, we get
\be
\a  = \frac{v^2/v_c^2}{v^2 - v_c^2} .	\label{agaandvvc}
\ee
In terms of these parameters, the transformations (\ref{StritoEsimetStritoEsidil}) 
read simply
\be
\begin{split}
g_{\m\n} &= \exp \big(  v_c^2  \Phi \big) G_{\m\n}
		= \exp \big(- v \ka \phi  \big) G_{\m\n} ,	
\\
\ka \phi &=  - ( v_c^2/ v) \,  \Phi .
\end{split}
		\label{PhiphiGgtrans}
\ee
Note that the constant appearing in the action (\ref{EuclStraga}) is found from (\ref{VdephiExpW}) to be 
\be
\begin{split}
C &= \frac{(d-1) (v_c/v)^4}{L^2} \Big[  d - (v/v_c)^2   \Big] 
\\
	&= \frac{( d - 2\a  ) (d-1 -2 \a  )}{\cal R^2}  ,
\end{split}\ee
where we have defined the radius
$\cal R \equiv \tfrac{2}{d-1} |\a | L$.

Now we can use  (\ref{PhiphiGgtrans}) to obtain solutions of (\ref{EuclStraga}) starting from the Einstein-frame domain wall solutions found in \S\ref{SectLiouvandAdS}.
From (\ref{phipfzExpsw}), we find the dilaton evolution
\bsub
\be
\exp   \Phi  = \left( \frac{|z_0 - z|}{\cal R}  \right)^{2 \a} ,
	\label{expPhizz0sol}
\ee
and then the metric $G_{\m\n}$ is found to be simply AdS space in Poincar\'e coordinates with radius $\cal R$,%
\footnote{%
This can be found by noting that
$$\omega^2 = e^{- v_c^2  \Phi} a^2(\Phi)
	= e^{-  \frac{v^2 - v_c^2}{v_2/v_c^2}   \Phi} 
	= e^{- \Phi / \a }
$$
where we have used Eq.(\ref{aofphexpW}), which gives $a^2 (\Phi) = \exp \big[  (v_c^4 /v^2)  \Phi \big]$.}
\be
\begin{split}
G_{\m\n} dx^\m dx^\n &= \omega^2(z) \big[ dz^2 + \eta_{ab} dx^a dx^b \big] ,
\\
\text{with}
\quad
\omega^2(z) &= \frac{\cal R^2}{(z_0 - z)^2}  .
\end{split}
	\label{AdSomeG}
\ee\label{LinDilaSolGPh}\esub
Eqs.(\ref{LinDilaSolGPh}) are an AdS linear dilaton solution of (\ref{EuclStraga}), i.e. the exponential of the dilaton evolves trivially as a power of the radial coordinate.

The existence of linear dilaton solutions in Liouville models allows for a precise definition of the holographic correspondence,  known as  `non-conformal holography' \cite{Boonstra:1998mp,Kanitscheider:2008kd}. Absence of conformal invariance is due to the (trivial) running of the dilaton/coupling, but the holographic QFT posses a `generalized conformal structure': the theory is invariant under Weyl transformations of the metric as long as the coupling is also appropriately transformed.

In the holographic dictionary, the scale factor $\omega$ in (\ref{AdSomeG}) is dual to the energy scale of the QFT.  
Holographic renormalization, first done in \cite{Kanitscheider:2008kd},  follows the same steps of standard holography  --- one makes a Fefferman-Graham (FG) expansion for the geometry, along with an expansion for the running dilaton,
\bsub\begin{align}
G_{\m\n}(\rho , x) &= \cal R^2 \frac{d\rho^2}{4\rho^2} + \frac{g_{ab}(\rho, x) dx^a dx^b}{\rho}
\\
\Phi(\rho , x) &= \a \log \rho + \kappa(\rho , x)
\end{align}
where $\rho = (z - z_0)^2/\cal R^2$ is the standard FG coordinate and the expansion has the form
\begin{align}
g_{ab}(\rho, x) &= g_{ab}^{(0)}(x) + \rho g_{ab}^{(2)}(x) + \cdots 
				+ \rho^\s \big[ g_{ab}^{(2\s)}(x) +\log \rho \,  h_{ab}^{(2\s)}(x) \big]
							+ \cdots
\\
\kappa(\rho,x) &= \kappa^{(0)}(x) + \rho \kappa^{(2)}(x) + \cdots 
				+ \rho^\s \big[ \kappa^{(2\s)}(x) + \log \rho \, \varphi^{(2\s)} (x) \big] + \cdots
		\label{FGexapnd}
\end{align}\label{FGexapn}\esub
The Einstein equations derived from (\ref{EuclStraga}) determine the functions $g^{(n)}_{ab}(x)$ and $\kappa^{(n)}(x)$ \emph{algebraically} in terms of the ``sources'' $g^{(0)}_{ab}(x)$ and $\kappa^{(0)}(x)$, up to order $n = \s$, where \cite{Kanitscheider:2008kd}
\be
\s = \tfrac{1}{2} d - \a  .
\ee
Such holographic reconstruction of the bulk metric (i.e. the coefficients $g^{(n)}_{ab}$ and $\kappa^{(n)}$) from the boundary structure (i.e. $g^{(0)}_{ab}$ and $\kappa^{(0)}$) is basically the same as in the pure AdS case \cite{Bianchi:2001de,Bianchi:2001kw,Papadimitriou:2004ap}, only now the order $\s$ of the non-local terms is shifted by $- \a$. 
From here, it is evident that we must have
\be
\a \leq d/2 ,	\label{bounaleqd2}
\ee
otherwise $\s < 0$ and the Fefferman-Graham expansion breaks down.
The most interesting cases have in fact $\a < 0$, even though for negative parameters with $|\a| < \frac{1}{2}$, the kinetic term in (\ref{EuclStraga}) is negative.
For example, the decoupling limit of D$p$-branes with $p = 0,1,2,3,4,6$, can be obtained from the action (\ref{EuclStraga}) if we set $d = p +1$, and 
$\a = \frac{(p-3)^2}{2(p-5)}$, see \cite{Kanitscheider:2008kd,Boonstra:1998mp}. 
All of these branes, except $p = 6$, have $\a \leq 0$.%
\footnote{%
Specifically, for $p = \{0,1,2,3,4,6\}$ we have 
$$\a = \{ - \tfrac{9}{10} , - \tfrac{1}{2} , - \tfrac{1}{6} , 0 , - \tfrac{1}{2} , \tfrac{9}{6} \}.$$}
The D6-brane has $\a = \frac{9}{2}$, violating (\ref{bounaleqd2}), hence there is no FG expansion since $\s = -1$. On the other hand,  a D2-brane has $\a = - \frac{1}{6}$, which gives a negative kinetic term in (\ref{EuclStraga}).

\bigskip

The effect of CSFI on the linear dilaton solutions is to change $\a$, as can be readily seen from Eq.(\ref{agaandvvc}) together with (\ref{qtqvtvtranf}), $\a \mapsto \tilde{\a} = - \a + 2/v_c^2$, i.e.
\be
 \tilde{\a} = - \a + (d-1).	\label{agaCSFImap}
\ee
Hence the image of (\ref{LinDilaSolGPh}) is again a linear dilaton AdS solution, but with a different exponent $\tilde {\a}$ for the evolution of $\exp \tilde \Phi$, and a different radius $\tilde{\cal R}$ for the AdS geometry.

We have seen that CSFI is only well-defined for parameters $v$ lying within the range (\ref{bongvvc}). The range splits into two halves, $v^2 \in [0 , v_c^2]$ and $v^2 \in [v_c^2 , 2 v_c^2]$, related by CSFI as in Fig.\ref{RangeOfExp}. 
These intervals translate into $\a$ as
\bsub\begin{alignat}{4}
&& v^2 &\in [0, \; v_c^2) \quad &&\longleftrightarrow \quad \a  &&\in [0 , \; - \infty)	\label{Intevaagvvca}
\\
&& \text{CSFI} &\Bigg\updownarrow 		&&{}	 	&&\Bigg\updownarrow \text{CSFI} \nonumber
\\
&& \tilde v^2 &\in (v_c^2, \; 2v_c^2] \quad &&\longleftrightarrow \quad \tilde \a  &&\in (\infty , \; d-1]	 	\label{Intevaagvvcb}
\end{alignat}\label{Intevaagvvc}\esub
Note the reversed orientation of the intervals in the r.h.s., due to the fact that, e.g.~in (\ref{Intevaagvvca}) we have $\{v^2 = 0\} \mapsto \{\a = 0\}$ and $\{v^2 = v_c^2\} \mapsto \{\a = - \infty\}$.
The boundedness of the interval of $v^2$ where CSFI is well defined translates into a gap $(0, d-1)$, for which $\tilde \a$ is not defined. 

There is a simple interpretation for the gap: \emph{iff $\a \notin (0, d-1)$ then $\a$ and $\tilde{\a}$ always have opposite signs.}
This can be immediately seen from the transformation (\ref{agaCSFImap}).   
Since $\a$ is the power appearing in the linear dilaton solution (\ref{expPhizz0sol}), this means after all that CSFI is a map such that
$\exp \Phi > 1  \Leftrightarrow  \exp \tilde \Phi < 1$.
Now, note that the positive values of $\a$ allowed by the bound (\ref{bounaleqd2}), viz. $0 < \a \leq \frac{d}{2}$, lie \emph{inside} the forbidden gap in (\ref{Intevaagvvc}), so these models do not have consistent CSFI images.
Meanwhile, any $\a \leq 0$ is mapped to a $\tilde{\a} \geq d-1$, thus violating the bound (\ref{bounaleqd2}), hence the image under CSFI of any model with $\a \leq 0$ (e.g. the $p$-brane solutions discussed) is a model for which the FG expansion (\ref{FGexapn}) is not defined.
But we can look the other way around: if we take a model in interval (\ref{Intevaagvvcb}), it is uniquely mapped by CSFI into a model with $\a \leq 0$, which can be consistently renormalized with the FG expansion (\ref{FGexapn}).
The special Liouville model has $\a = d-1$ and is mapped to $\a = 0$,  which is (like in the Einstein frame) pure AdS since the dilaton (\ref{expPhizz0sol}) freezes. (One must redefine $\Phi \to \a \Phi$ in the action before taking $\a = 0$.)

From the discussion above, one can already conclude that the only effect of CSFI in the linear dilaton solutions is to change the exponent of the dilatonic evolution: the
 Jordan-frame scale factor $\omega$ rests \emph{invariant}. Let us show this explicitly. 
Using (\ref{PhiphiGgtrans}) and (\ref{aofphexpW}), we find that
\be
\omega(z) = \left[ a(z) \right]^{\frac{v_c^2 - v^2}{v_c^2}}.	\label{omegaaz}
\ee
Making the CSFI transformation, we have to take two effects into account: first, that the Einstein-frame scale factor $a$ undergoes an inversion, second that the exponent is transformed according to (\ref{qtqvtvtranf}). The overall effect is therefore
\begin{equation*}
\begin{split}
\tilde \omega(\tilde z) &= \left[ \tilde a (\tilde z) \right]^{\frac{v_c^2 - \tilde v^2}{v_c^2}} 
	= \Big[ \frac{1}{a(z_* \pm z)} \Big]^{\frac{v^2 - v_c^2}{v_c^2}}
	=   \left[ a(z_* \pm z) \right]^{\frac{v_c^2 - v^2}{v_c^2}}
\\	
&=   \omega(z_* \pm z) .
\end{split}	
\end{equation*}
For simplicity, we have set $\c = 1$ in (\ref{sfacinvers}), by  the usual freedom of choosing the normalization of the scale factor. We recall that the transformation of the argument, $\tilde z = z_* \pm z$ is also the usual translation- and reflection-invariance of the domain wall solutions.

Finally, let us make a comment about the invariant potential in the Einstein frame, with $v = v_c$. It corresponds to $\a = \pm \infty$, then the kinetic term of $\Phi$ is uniquely normalized to unity in (\ref{EuclStraga}) and, with the redefinition $\Phi \to -2 \Phi$, one finds just the bosonic string effective action, but with a constant dilaton potential 
\be
\begin{split}
S &=  \int \!\! d^{d+1} x \, \sqrt{-G}  \frac{e^{-2 \Phi}}{\ka^2}  \Big[ \scr R  + 4 G^{\m\n} \pa_\m \Phi \pa_\n \Phi - \frac{(d-1)^2}{L^2} \Big] 
\\
 (v &= v_c) .
\end{split} 
\label{footnoteSframe} 
\ee
Again, there is no meaningful FG expansion, since $\s \to \pm \infty$. This was to be expected, as the solution of the Liouville model is qualitatively different in this case. 
The Einstein-frame solution (\ref{SolforazExp2}) and (\ref{phipfzExpswCr}) is \emph{not} an AdS linear dilaton solution in the Jordan frame. 
In fact, from Eq.(\ref{omegaaz}) we can see that in the critical model the Jordan-frame scale factor is trivial, $\omega = 1$. The Jordan-frame geometry is, therefore, simply Minkowski space, 
while the dilaton, given by (\ref{PhiphiGgtrans}) and (\ref{phipfzExpswCr}), is
$\Phi = \frac{(d-1)}{L} (z_0 - z)$.

\subsubsection{CSFI and dimensional reductions of AdS}	\label{SectNonConfpBrn2}

We have seen that Einstein-frame Liouville models with parameters lying within the range (\ref{Intevaagvvca}) have a well-defined holographic interpretation via the AdS linear dilaton solution (\ref{LinDilaSolGPh}) found in the Jordan frame (\ref{EuclStraga}).
In \cite{Kanitscheider:2009as}, it was shown that one can obtain the Einstein-frame potentials in a different way: by performing a dimensional reduction of a higher-dimensional pure AdS solution over a torus. 
The reduction is consistent, hence one can first perform the usual renormalization of the $D$-dimensional asymptotically AdS solution, with a FG expansion (\ref{FGexapn}) of order $\s = \frac{1}{2} (D-1)$; then reducing over a torus of dimension $q = 2\s - d$  yields the correct counterterms and sourced one-point functions for the non-conformal holography of the $(d+1)$-dimensional Liouville models. 

Later \cite{Gouteraux:2011qh}, it was shown that the dimensional reduction can be done consistently not only over a torus but over any Einstein manifold.  One starts from a pure Einstein action in $D=2\s+1$ dimensions,
\be
S = \frac{1}{\ka^2_{D}} \int d^D x \ \sqrt{- G_{(D)}} \, \big( R^{(D)} - 2 \La_{D} \big) ,
 \quad D = 2\s + 1 ,
	\label{Act2sigDim}
\ee
and performs a  reduction to a $(d+1)$-dimensional theory by factoring out an internal space $X^q$ with dimension $q$, 
\be
ds^2_{(D)} = e^{2 \la \phi} ds^2_{(d+1)} + e^{2 \beta \phi} ds^2_q(X^q) ,
\quad 
q = 2\s -d
.
	\label{DimsReducAns}
\ee
The reduction is consistent, i.e. it gives the correct Einstein equations in lower dimensions, if the parameters are related by  $(d-1) \la = - q \beta$, and if 
 $X^q$ is an Einstein manifold (which we assume to be compact), i.e. its Ricci curvature must be proportional to its metric, with curvature normalization
$R^{(q)}_{ij} = (1/q) R^{(q)} g^{(q)}_{ij}$.
Then (\ref{Act2sigDim}) becomes the standard Einstein-Hilbert action, with the Kaluza-Klein field $\phi$ appearing as a scalar with  potential
\be
V(\phi) = \frac{2 \La_D}{\ka^2} e^{- v_q \ka \phi} - \frac{R^{(q)}}{\ka^2} e^{- w_q \ka \phi} ,
	\label{expVdimreduc}
\ee
where
\be
\begin{split}
v_q &= v_c \, \sqrt{\frac{2\s -d}{2\s - 1}} 
	=  \frac{v_c^2}{\sqrt{v_c^2 + 2/q}}  ,
\\
w_q &=  v_c \, \sqrt{\frac{2\s - 1}{2\s -d} } 
	=  \sqrt{v_c^2 + 2/q} .
\end{split}	
 	\label{vswsexpn}
\ee
Recall that $v_c = \sqrt{2/(d-1)}$.
Canonical normalization of $\phi$ implies that $\la = - \frac{1}{2} v_\s$.
Newton's constant  is induced from (\ref{Act2sigDim}) as $\ka^2 = \ka^2_D / V_q$, where $V_q$ is the volume of the Einstein space $X^q$.

The dimensional reduction (\ref{DimsReducAns}) can be `generalized' by noting that $q$ only enters the low-dimensional physics as a parameter in (\ref{expVdimreduc}) \cite{Kanitscheider:2009as}; then one can continue $q$ (and $\s$) away from a (semi-)integer.
Thus $v_q$ becomes a continuous parameter, but still it must satisfy 
\be
0 \leq v_q^2 \leq v_c^2 
\quad
\text{because}
\quad
q > 0
, 
\label{boundsdim1}
\ee
that is the (generalized) dimension of the compactified space $X^q$ must be non-negative.
Meanwhile, 
\be
 v_c^2 \leq w_q^2 \leq d \ v_c^2 \quad
\text{because}
\quad
q \geq 1
. \label{boundsdim2}
\ee
The upper bound coincides with the condition (\ref{vvmaxbonud}) that the potential be negative, but here it has a very different interpretation: it amounts to the dimension of $X^{q}$ not being smaller than one.

\bigskip

The parameters $v_q$ and $w_q$ lie in the complementary ranges (\ref{SFDwsaprangv}) related by CSFI in single-exponential Liouville models. 
There are two instances when only one of the two exponentials in the sum (\ref{expVdimreduc}) is in fact present. For each case, the $V_0$ in (\ref{VxpVexpSF}) has a different interpretation:

\begin{description}[font={\normalfont\itshape}]

\item[A. Higher-dimensional cosmological constant]
If the internal space has zero curvature, i.e. if $X^q$ is a torus, the potential (\ref{expVdimreduc}) has one single exponential whose exponent lies in the range (\ref{boundsdim1}). Then $V_0 = 2 \La_D / \ka^2$  is inherited from the cosmological constant of the higher-dimensional AdS space.

\item[B. Curvature of internal space]
Instead, if $\La_D = 0$ and the internal space has positive curvature $R^{(q)} > 0$, i.e. $X^q$ is a sphere, the (negative) potential will have one single exponential with exponent in the range (\ref{boundsdim2}). In this case, $\tilde V_0 = - R^{(q)} / \ka^2$ comes from the curvature of $X^q$.

\end{description}
So in cases A and B, the single exponentials in $V(\phi)$ have exponents in the ranges (\ref{boundsdim1}) and (\ref{boundsdim2}), respectively. CSFI swaps these two complementary intervals because of (\ref{qtqvtvtranf}) (cf. Fig.\ref{RangeOfExp}), and therefore it also swaps the two interpretations: the image of a type A potential is a type B potential, and vice-versa. 
In short, CSFI maps $v_q \leftrightarrow w_q$ and $\rm{A} \leftrightarrow \rm{B}$. But there is a subtlety: the relation between $v_q^2$ and $w_q^2$ that follows from the dimensional reduction formulae (\ref{vswsexpn}) is
\be
w_q^2 = v_c^4 / v_q^2 ,	\label{vswsexpnAg}
\ee
which is, of course, not the same as (\ref{qtqvtvtranf}).
Hence the pair of potentials related by CSFI correspond not only to different ``sectors'' of the dimensional reduction, i.e. to either $\La_D$ or $R^{(q)}$ being zero, but they also 
correspond to the reduction of a \emph{different number of dimensions}, i.e. to different values of $q$.
Precisely, identifying $v = v_q$ and $\tilde v = w_{\tilde q}$ and using Eqs.(\ref{vswsexpnAg}) and (\ref{qtqvtvtranf}), we have 
\begin{equation*}
\tilde v^2  = v_c^2 + \frac{2}{\tilde q} 
		= 2 v_c^2 - v^2 
		= 2 v_c^2 - \frac{v_c^4}{v_c^2 + 2/q}  
\end{equation*}
and solving for $\tilde q$,
\be
\tilde q = q + (d-1) .
\ee

Let us summarize this geometric interpretation. 
CSFI relates a pair of Liouville models, \emph{both} in $(d+1)$ dimensions. 
One of the models, type A, is obtained from dimensional reduction over a torus $T^q$ of a pure $AdS_D$ solution in $D = (d+1) + q$ dimensions. 
The other model in the pair, type B, is obtained by dimensional reduction over a sphere $S^{\tilde q}$ of a \emph{flat} solution in $\tilde D = (d+1) + \tilde q$ dimensions.
The generalized (possibly continuous) dimensions $q$ and $\tilde q$ of the internal Einstein spaces, and the dimensions $D$ and $\tilde D$ of the total reduced spacetime are related by
\be
\tilde q - q = \tilde D - D = d - 1 .		\label{qtlqDtiD}
\ee
This relation is asymmetric for tilded and untilded quantities, as it was expected since now we have different properties for each of the solutions in the CSFI pair. 
In particular, Eq.(\ref{qtlqDtiD}) only makes sense if both $q$ and $\tilde q$ are positive, hence
\be
\tilde q \geq d-1 ,	\label{tileqbounCSFO}
\ee
which is precisely equivalent to the bound $0 \leq w_q^2 \leq 2 v_c^2$ in (\ref{bongvvc}).
(Note that this CSFI bound is stricter than (\ref{boundsdim2}).)

By this construction, the $AdS_{d+1}$ solution obtained from the Liouville models by making $v = 0$ corresponds to a model of type A with \emph{no} dimensional reduction, i.e. $q = 0$ hence $D = d+1$. CSFI maps it to the special Liouville model, which is of type B with the dimension of the reduced sphere $\tilde q = d -1$, saturating the bound (\ref{tileqbounCSFO}); in this special case, $\tilde D = 2d$. As for the self-invariant Liouville model, with $v_q = w_q = v_c$, it can only be achieved in the limit of \emph{infinite} dimensions: i.e. $q, \tilde q \to \infty$ and $D , \tilde D \to \infty$ with fixed ratios $q / \tilde q = D / \tilde D = 1$.

We have seen in \S\ref{SectNonConfpBrn} that Liouville models with $\tilde v \in (v_c^2 , 2 v_c^2)$ correspond to linear dilaton solution with an ill-defined FG expansion. 
Now we have seen that these models lift to a $D$-dimensional asymptotically \emph{flat} spacetime where, indeed, there is no FG expansion.%
\footnote{%
A note: in making the FG expansion directly in the $(d+1)$-dimensional space, as in \S\ref{SectNonConfpBrn}, the failure of the expansion was due to having $\s < 0$. In the higher-dimensional construction, $\s$ is always positive, and the absence of the FG expansion is due to a different reason, namely the $D$-dimensional spacetime not being asymptotically AdS.}
CSFI maps these models to their well-behaved, asymptotically AdS partners.

\subsection{Asymptotically AdS solutions and their pairs}	\label{SectAsumptAdSanddual}

Now consider an asymptotically AdS domain wall  $\DW$. 
The superpotential has the standard form corresponding the quadratic approximation of the potential near a maximum,
\begin{align}
& W (\phi) =  \frac{d-1}{\ka \ell} + \frac{s }{4\ka \ell} (\ka \phi)^2 + \cdots 
	\label{SuperCJs}
\\
& V(\phi) = - \frac{d(d-1)}{\ka^2 \ell^2} + \tfrac{1}{2} m^2 \ \phi^2  + \cdots ,
	\label{AdSVnearquaphi}
\\
&	s = \tfrac{1}{2} \big( d \pm \sqrt{d^2 + 4 m^2 \ell^2} \big) > 0 ,
\quad 
	- \tfrac{d^2}{4\ell^2} < m^2 < 0 .
\end{align}
We are considering solutions with $0 < \ka \phi  \ll 1$, near the AdS vacuum with radius $\ell$ located at $\phi = 0$.
 The geometry will be the vicinity of the AdS boundary.
The function (\ref{betaM}) is 
\be
 \beta(\phi) = s  \phi +\cdots
\label{betwxps}	
\ee
To find $\tilde \phi$, we must insert this into Eq.(\ref{sigma_dual2M}) and integrate. Keeping only the leading term,
\be
\ka\tilde \phi = - \frac{2 \sqrt{d-1}}{s}  \log (\ka \phi) + \cdots 
	\label{Integrexptidlphph}
\ee
For $0 < \ka \phi \ll 1$, we have $\ka \tilde \phi  \gg 1$. Inverting (\ref{Integrexptidlphph}) as $\phi = \phi(\tilde \phi)$, 
\be
\ka \phi = \Big(  e^{- \frac{\ka\tilde \phi}{2 \sqrt{d-1}} } \Big)^s \left[ 1 + \rm{O} \Big( e^{- 2s \frac{\ka\tilde \phi}{2 \sqrt{d-1}} } \Big) \right] 
	\label{phioftlaint}
\ee
where $e^{ - \frac{\ka\tilde \phi}{2 \sqrt{d-1}} }  \ll 1$.

\begin{figure}
    \begin{center} 
    \begin{subfigure}
        \centering
        \includegraphics[height=1in]{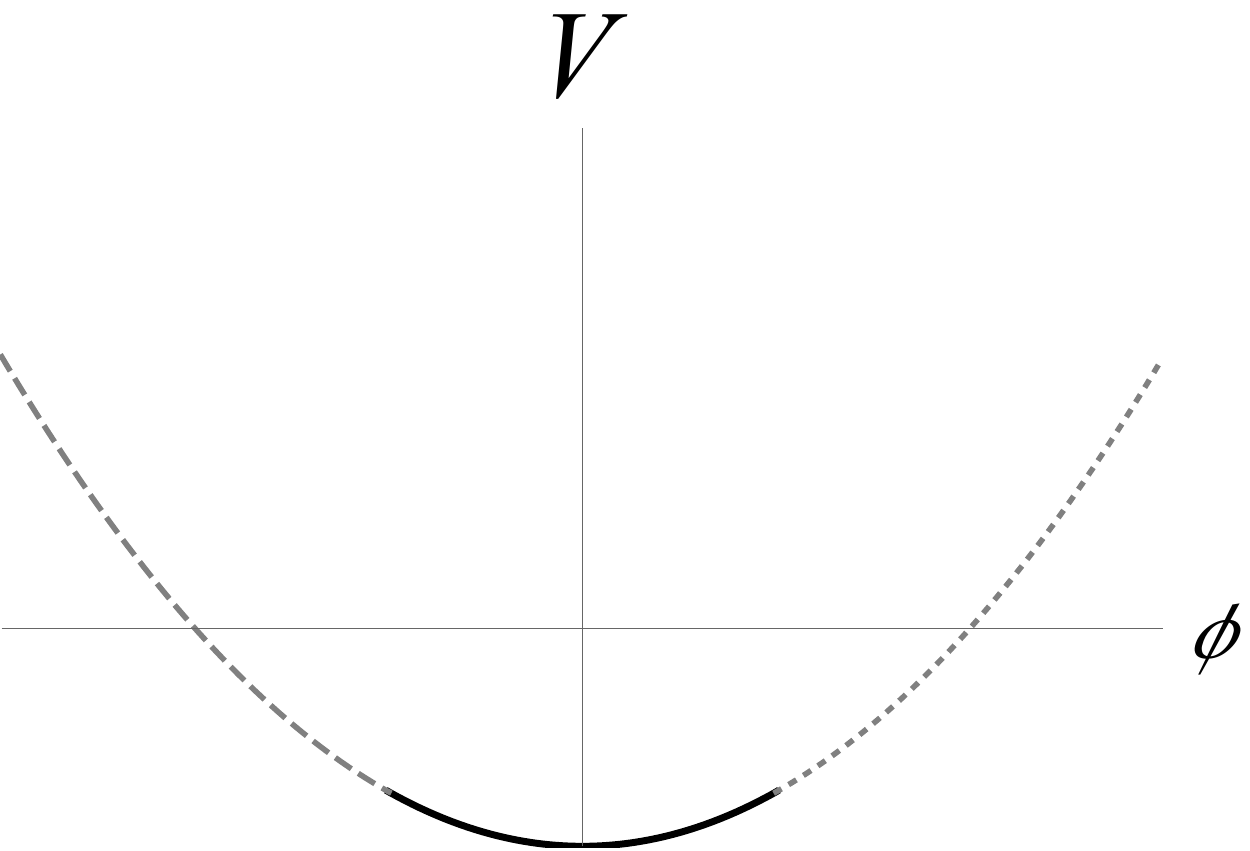}
    \end{subfigure}%
\qquad
    \begin{subfigure}
        \centering
        \includegraphics[height=1in]{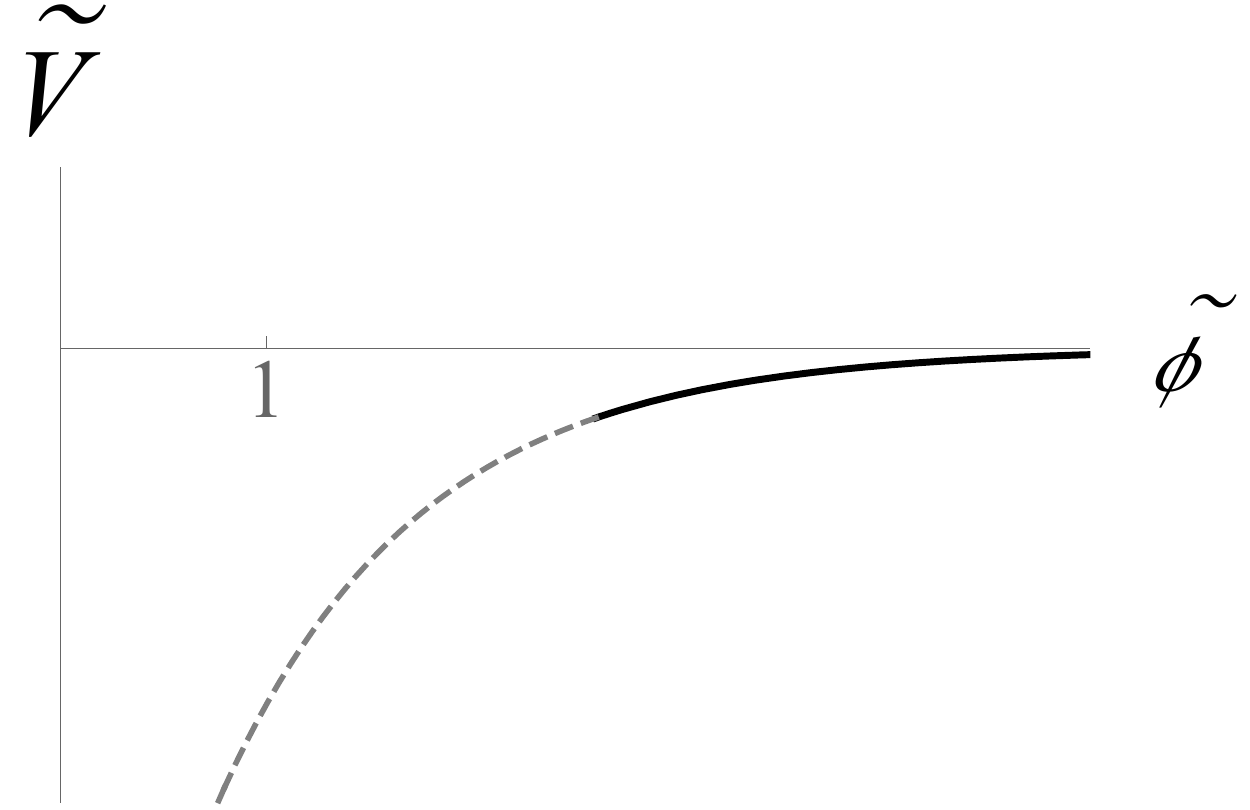}
    \end{subfigure}
    \end{center}
    \caption{Asymptotic regions of related potentials.
    (a) The quadratic potential (\ref{AdSVnearquaphi});
    (b) The Liouville potential (\ref{VphnearSingSFDex})
    }
    \label{dualVtiV}
\end{figure}

We now want to find the superpotential for $\tDW$. Inserting (\ref{betwxps}) into Eq.(\ref{dualbetapm}) and using (\ref{phioftlaint}) we get
\be\begin{split}
 \tilde \beta(\tilde \phi) =  \frac{2 \sqrt{d-1}}{\ka} &- \frac{s^2}{4 \sqrt{d-1} \ka} \Big(  e^{- \frac{\ka\tilde \phi}{2 \sqrt{d-1}} } \Big)^{2s}
  	+ \rm{O} \Big(  e^{- 4s \frac{\ka\tilde \phi}{2 \sqrt{d-1}} } \Big) .
\end{split}	\label{tildebetalas}
\ee
Eq.(\ref{betaM}) can be solved as a differential equation for $\tilde W (\tilde \phi)$, resulting in an exponential superpotential, and in the asymptotic expressions
\begin{align}
& \tilde W (\tilde \phi) = \tilde W_0 \, e^{ \frac{\ka \tilde \phi}{ \sqrt{d-1}}}  
		\Bigg[ 1 + \tfrac{s}{8(d-1)}  e^{- s \frac{ \ka \tilde \phi }{ \sqrt{d-1}}} 	+ \cdots \Bigg] 
\\
& \tilde V ( \tilde \phi) = -  \tfrac{d-2}{d-1}  \tilde W_0^2 e^{ \frac{2 \ka \tilde \phi}{\sqrt{d-1}} } \Bigg[ 1 
		+ \tfrac{s (2s + d -2)}{4(d-1)(d-2)} \ e^{ - s  \frac{ \ka \tilde \phi}{\sqrt{d-1}} }
			 \Bigg]
			\label{VphnearSingSFDex}
\end{align}
with $ \ka \tilde \phi \gg 1$. 
The first term of $\tilde V(\tilde \phi)$ is the leading one, and it is just the special-Liouville potential; the other terms vanish for $\tilde \phi \to \infty$. We plot asymptotic regions of the pair of potentials (\ref{AdSVnearquaphi}) and (\ref{VphnearSingSFDex}) in Fig.\ref{dualVtiV}.

Of course, if we include more (subleading) terms in the ellipsis in Eq.(\ref{SuperCJs}), then the formula (\ref{VphnearSingSFDex}) for  $\tilde V(\tilde \phi)$ will include more terms inside the brackets. 
But note that the leading term does \emph{not} depend on $s$; so we could, for example, start with $s = 0$ and get the same potential at leading order. Say, if instead of (\ref{SuperCJs}) we start with a \emph{cubic} superpotential,
\[
W(\phi)\approx \frac{d-1}{\ka \ell}+\frac{\mathcal{D}}{6\ka \ell}(\ka\phi)^3, \qquad \ka\phi\ll1,
\]
the domain wall structure is very different, but after calculating $\tilde W (\tilde \phi)$ and $\tilde V(\tilde \phi)$ we still find the same leading asymptotic exponential behavior.
This was expected, since AdS is always mapped by CSFI to the special Liouville solution.

\subsection{Invariant domain walls}	\label{SectSelfdual}

An obviously interesting class of solutions are geometries invariant under the scale factor inversion,
\be
a(z) \mapsto \tilde a( z) = a(\tilde z) ,
\quad
 \text{hence}
\quad
a(z) = \frac{\c^2}{a(2z_c - z)} .
\ee
The position $z_c$ is the invariant point of the transformation, where $\tilde a(z_c) = a(z_c) = \c$.
In \cite{dS:2015fwa}, it was shown how to construct invariant cosmological solutions, by making an appropriate ansatz for the energy-density scaling with $a$. We now adapt the argument for domain wall solutions. 

The first step is to consider an ansatz for the superpotential as a function of the scale factor, i.e. $W = W(a)$, and then impose the condition of invariance, which from Eq.(\ref{WatsnfSFD}) reads
\be
W^2(\tilde a) = (a / \c)^4 W^2(a) .
\ee
A simple non-trivial ansatz is
$W^2(a) = ( w_1 a^r + w_2 a^s )^t$, whence invariance implies that $t = -2/(r+s)$ and relates  $w_2/w_1 = \c^{r-s} $. 
We will focus in models with $r = 0$ which, after some relabeling, can be written as
\be
W(a) = W_0 [ 1 + (\c / a)^{4 \a} ]^{\frac{1}{2\a}} . \label{SfDuWa}
\ee
These are the models which are asymptotically AdS.

Once given $W(a)$, one must solve the first-order equations to find $W(\phi)$. In general this is hard, but for  (\ref{SfDuWa}) it can be easily done. We are required to find $a = a(\phi)$; to do so, we must solve the equation  Eq.(\ref{betaM}) in the form
\be
da / d\phi = - a / \beta ,
\quad
\text{hence}
\quad
\phi = - \int \frac{da}{a\beta(a)} .
\ee
We can manipulate Eq.(\ref{betaM}) to write $\beta$ as a function of the scale factor,
\be
\beta^2(a) = - \frac{2(d-1)}{\ka^2} \frac{a \ d W/da}{W(a)} ,
\ee
and inserting (\ref{SfDuWa}) into these equations, we find that
\be
\ka \phi = - \frac{\sqrt{d-1}}{\a} \ \rm{arc \, coth} \left[ \sqrt{1 + (w/\tilde w)a^{4\a}} \right] ,
\ee
therefore
\be
W(\phi) = W_0 \left[ \cosh \left( \frac{\a}{\sqrt{d-1}} \ka \phi \right) \right]^{\frac{1}{\a}} .
	\label{SfDuWphi}
\ee
This is a class of invariant superpotentials parameterized by $\a$, hence a class of models for which CSFI is a \emph{symmetry} of the solutions.
In the next section, we show that the model with $\a = 1/2$ is the well-known GPPZ flow.

\section{Holographic implications of CSFI}	\label{SectHoloImpl}

The first-order system (\ref{EinsEqHombothZ}) has a standard holographic interpretation as  RG equations of a QFT$_d$ with an energy scale $E = a(z)$ and running coupling $g = \phi(z)$, driven by a scalar operator $\scr O$ \cite{deBoer:1999tgo,Bianchi:2001de}.  The RG  flow is determined by the holographic beta-function $\beta \equiv - dg/d \log E$ given by Eq.(\ref{betaM}). 
In this context,  CSFI defines a map between the  RG flows of two holographic QFT$_d$s --- a map which, because of scale factor inversion, relates their ultraviolet and infrared limits.
In \S\ref{SectSFDasdulflows} we discuss some properties of pairs of holographic RG flows, concentrating on the transformation of the beta-functions. 
In \S\ref{SectRenorActio}, we digress about the effect of CSFI on the one-point function of the operator $\scr O$.

\subsection{A relation between holographic RG flows}  \label{SectSFDasdulflows}

Let us call $\QFT$ and $\tQFT$ a pair of theories related by the CSFI correspondence.
The transformation of the beta-functions, given by Eq.(\ref{dualM}) (and using (\ref{sigma_dual2M})), 
\be
\beta^2 (\phi) + \tilde \beta^2 (\tilde \phi ) = 4(d-1)/ \ka^2 ,
						\label{dualM2}
\ee
now should be seen as a statement about the RG flows; the RG flow of $\QFT$ determines the RG flow of $\tQFT$.
Other holographic features are inherited by the domain wall solutions and the corresponding action of CSFI.
For example, the running of the anomalous scaling dimension of $\scr O$, given by $s(\phi) = - d \beta / d \phi$, transforms as%
\footnote{%
To show this, just take the derivative of Eq.(\ref{dualM2}), and use the identity $d \phi / \beta(\phi) = - d \tilde \phi / \tilde \beta(\tilde \phi)$
obtained from the definition of $\beta$ and the fact that scale-factor inversion gives $d a / a = - d \tilde a / \tilde a$.}
\begin{equation*}
\tilde \beta^2(\tilde \phi) \tilde s ( \tilde \phi) = \beta^2(\phi)  s ( \phi ).
\end{equation*}
Another important characteristic of the QFTs is the central-charge function%
\footnote{In Einstein gravity, and in asymptotically AdS$_{d+1}$ backgrounds, the two central functions ${\mathfrak c} $ and ${\mathfrak a}$ coincide: ${\mathfrak c} = {\mathfrak a}$ \cite{Henningson:1998gx}.}
\cite{Girardello:1998pd,Freedman:1999gp,Henningson:1998gx}
whose CSFI transformation is found readily from Eq.(\ref{WatsnfSFD}),
\begin{equation*}
{\mathfrak c} (\phi) \equiv \frac{2 (d-1)^{d-1} \pi^{d/2}}{\ka^{d+1} \Gamma(\tfrac{d}{2}) \left[ -W(\phi) \right]^{d-1}} \ ,
\quad
\frac{\tilde {\mathfrak c} (\tilde \phi)}{\tilde a^{d-1}(\tilde \phi)} = \frac{{\mathfrak c}(\phi)}{a^{d-1}(\phi)} .
\end{equation*}
At the UV fixed point of the RG flow (i.e. at  the AdS$_{d}$ boundary of the DW bulk), $\frak c (=\frak a)$ is a natural  generalization of the central charge(s) of the corresponding  UV CFT$_d$s.  In RG flows between two fixed points, this function obeys the (generalized) Zamolodchikov's $c$-theorem: $\frak c_\UV \geq \frak c_\IR$, assuring  the decreasing of the central functions. The inequality follows if the null energy condition is satisfied along the domain wall evolution \cite{Freedman:1999gp};
under the condition (\ref{dualbetabound}) CSFI preserves the NEC, so it also guarantees the validity of the $c$-theorem, i.e. $\tilde{\frak c}_\UV \geq \tilde{\frak c}_\IR$.

%
\begin{figure}[t]
\centering
\includegraphics[scale=0.43]{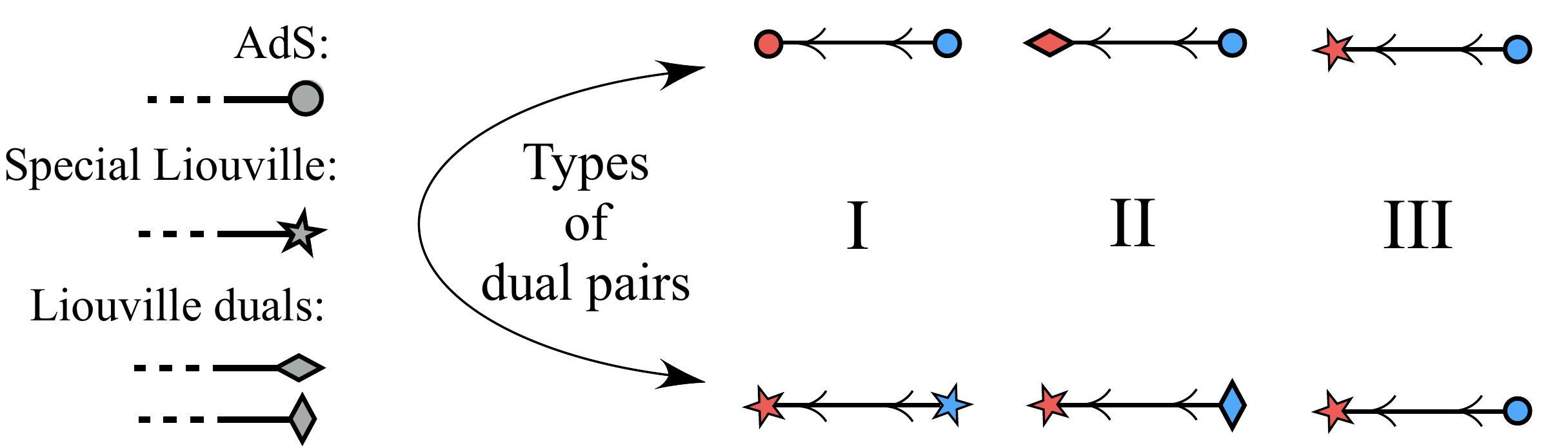}
\caption{Three types of RG flows related by CSFI. Bullets indicate AdS fixed points, stars indicate special Liouville asymptotics and the pair of spades indicate pairs of Liouville asymtptotics. Blue marks the UV and red the IR.}
\label{TypesOfDuals}
\end{figure} 
%

The map (\ref{dualM2}) does not preserve the RG fixed points, i.e.  
$\{ \beta(\phi) = 0 \} \mapsto \{\tilde \beta(\tilde \phi) \neq 0 \}$. 
Instead, as seen in \S\ref{SectAsumptAdSanddual}, CSFI maps $\QFT$ near the UV (AdS) fixed point with small $\ka\phi \ll 1$ to the IR regime of $\tQFT$ with a logarithmically diverging coupling $\ka \tilde \phi \gg 1$.
Such ``flows to infinity'' of the scalar field in the IR are relevant in phenomenological applications of holography to QCD \cite{Gursoy:2007cb,Gursoy:2007er,Gursoy:2016ggq,Gubser:2008yx}, and in holography applied to condensed matter \cite{Gouteraux:2011ce, Gouteraux:2011qh}.
For example, using the methods of \cite{Gursoy:2007er}, we can show that the special-Liouville singularity is ``confining'', in the sense that the holographic Wilson loops of the QFT obey the area-law  in the IR. Flows that end in a naked singularity in the IR are an indication of a non-trivial IR structure of the holographic QFT and, on this note, all our Liouville singularities are `good' by the criterion of Gubser \cite{Gubser:2000nd}, since the (negative) potential $V(\phi)$ is bounded from above. This means that the singular domain walls can be obtained from a black hole solution whose horizon shrinks to zero, and the field theories have a well-defined finite-temperature limit. 

As an illustration of what kinds of pairs of RG flows can be obtained  as a result of the CSFI correspondence, we now consider three examples of flows with a UV fixed point, and their respective images shown in  Fig.\ref{TypesOfDuals}.

\begin{enumerate}[\bfseries I)]

\item
In the standard RG flow between two CFTs the domain wall interpolates between two AdS vacua where $V(\phi)$ has a maximum (the UV point) and a minimum (the IR point), with $s (\phi_\IR) < 0 < s(\phi_\UV) $. The image of this flow interpolates between two special-Liouville asymptotics. 

\item
The flow starts in the AdS UV and runs to a general Louville singularity in the IR. Its image has two different Liouville asymptotics, with the special potential in the IR limit (as image of the AdS boundary). 

\item
The third example is a flow starting at AdS and ending with a special Liouville IR; its image has, therefore, the same types of asymptotics.

\end{enumerate}

\begin{figure*}
     \centering
        \includegraphics[scale=0.35]{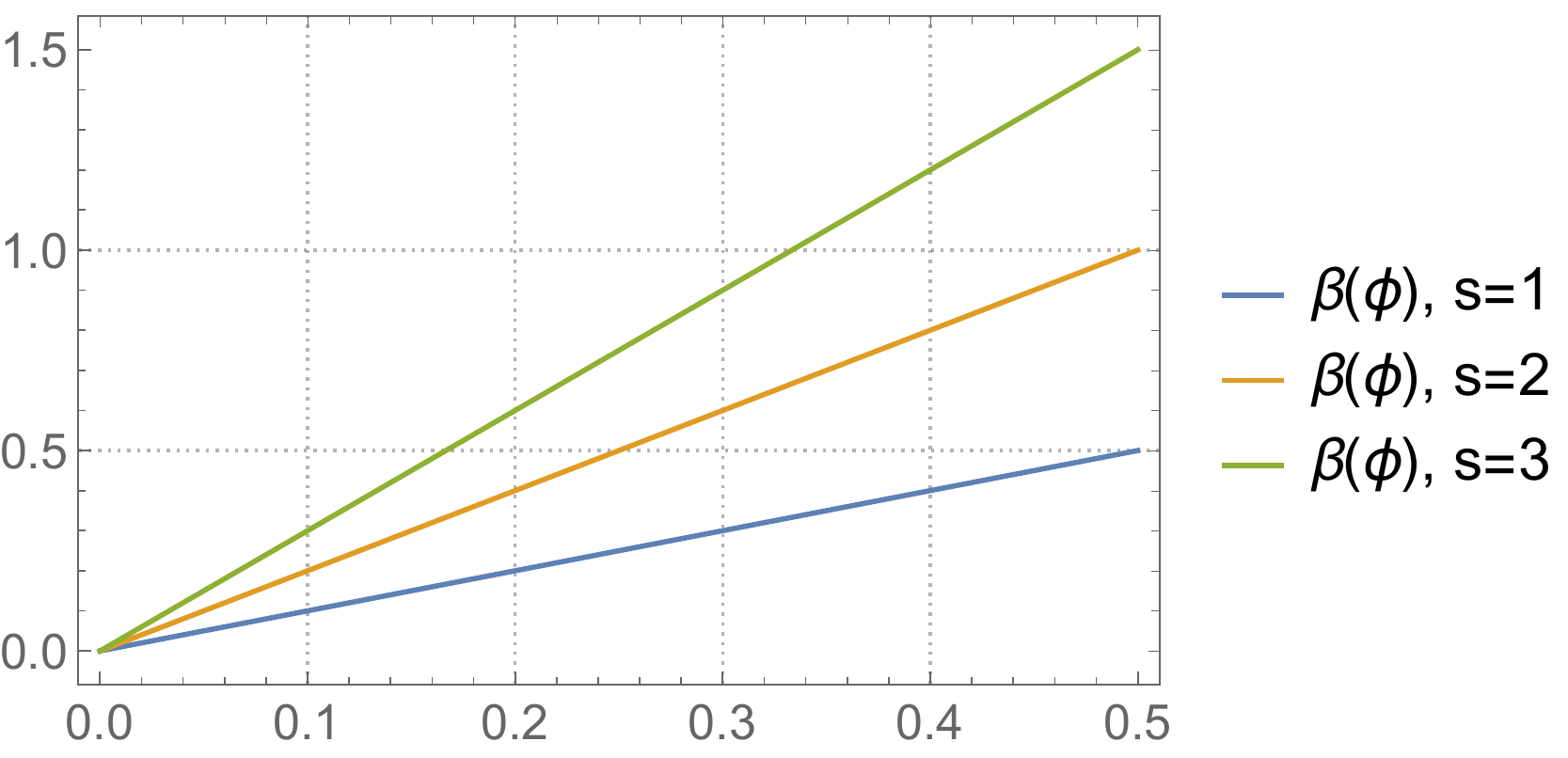}
        \includegraphics[scale=0.35]{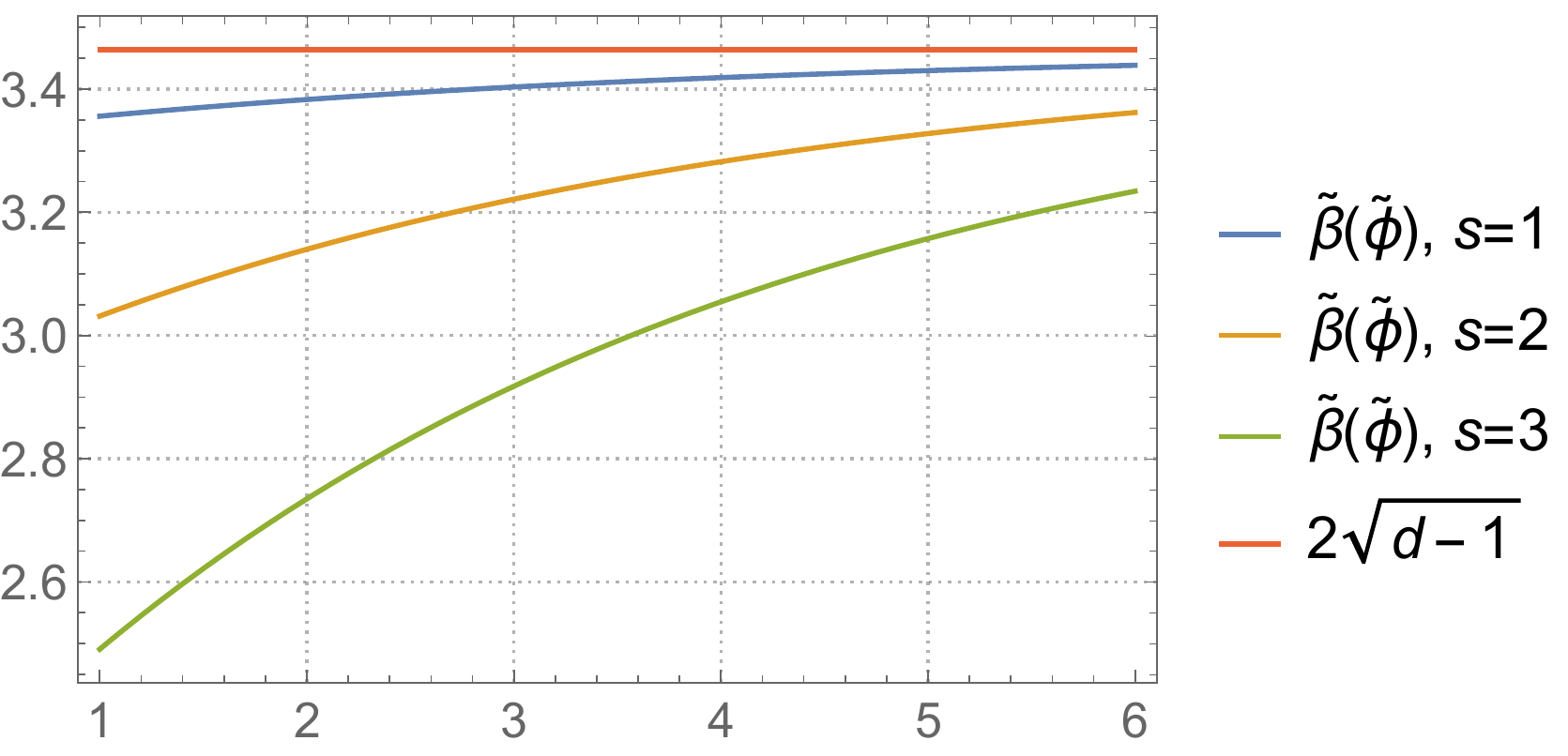}    
    \caption{%
    $\beta(\phi)$ near an (AdS) UV fixed point (left panel) and $\tilde \beta(\tilde \phi)$ near a Liouville singularity (right panel).
    }
    \label{dualbetastbat}
\end{figure*}

In \S\ref{SectAsumptAdSanddual} we have given the explicit expression of $\tilde \beta(\tilde \phi)$ and $\tilde \phi (\phi)$, mapped to the vicinity of a UV fixed point of $\beta(\phi)$ when $0 \ll \ka \phi \ll 1$. We saw that $\tilde \phi \approx - \frac{2 \sqrt{d-1}}{\ka s} \log (\ka \phi)$, so as $\phi \to 0^+$ we have $\tilde \phi \to + \infty$. The beta-functions $\beta(\phi)$ and $\tilde \beta(\tilde \phi)$, given in Eqs.(\ref{betwxps}) and (\ref{tildebetalas}), are plotted in Fig.\ref{dualbetastbat}.

In general, given a beta function $\beta(\phi)$, it is not possible to invert Eq.(\ref{sigma_dual2M}) to find $\phi ( \tilde \phi)$, hence we cannot give a closed expression for $\tilde \beta(\tilde \phi)$. However, it \emph{is} possible to give $\tilde \beta(\phi)$, so we can still see the behavior of the RG flow of $\tQFT$, even without knowing the explicit transformation of the fields. 
Let us illustrate this with examples of the three types above.

The simplest example of  Type I flow is given by the superpotential
\be
\begin{split}
W(\phi) = \frac{1}{\ka \ell} \left[ 2 + \a^{-2} s \sin^2( \tfrac{\a}{2} \ka \phi) \right]
 ,
\quad
\ka \beta(\phi) = \frac{s \sin( \a \ka \phi)}{1 + \frac{s}{3\a} \sin^2 (\frac{1}{2} \a\ka\phi)} .
\end{split}
\ee
We take $d = 4$.
There is a UV fixed point at $\phi_\UV = 0$ with anomalous dimension $s(\phi_\UV) > 0$, and a IR fixed point at $\phi_\IR = \pi / \ka \a$ with $s(\phi_\IR) = - 3\a^2 s / (3\a^2 + s) < 0$.
It pair $\tilde \beta$, as a function of the \emph{original} field $\phi$, can be readily found from Eq. (\ref{dualbetapm}).

A well-studied flow of Type II is the Coulomb branch of $\cal N = 4$ SYM in $d = 4$ . The superpotential is given by \cite{Bianchi:2001kw,Bianchi:2001de}
\be
\begin{split}
W(\phi) = \frac{2}{\ka} \Big[ e^{- \ka \phi / \sqrt6} + \tfrac{1}{2} e^{2 \ka \phi / \sqrt6} \Big] \ ,
\quad
\beta(\phi) = \frac{2\sqrt6}{\ka} \left( \frac{e^{\sqrt{3/2} \, \ka \phi} - 1}{e^{\sqrt{3/2} \, \ka \phi} + 2} \right) .
\end{split}
\ee
The RG flows from an AdS UV fixed point at $\phi = 0$, driven by the VEV of an operator with anomalous dimension $s = 2$, up to a null Liouville singularity at $\phi = - \infty$ (produced by a disc of D3-branes in the ``lifted'' 10-dimensional SUGRA), where  $W \sim  e^{-  v_c \ka \phi / 2}$. Here $v_c = \sqrt{2 / 3}$ is the critical parameter (\ref{VdephiExpW}).
Note that the critical-Liouville limit is invariant, as expected. The image of the UV point has $\tilde \beta \approx - 2\sqrt{d-1} / \ka$; the minus sign comes from the fact that $\beta$ and $\phi$ are negative.
Here we can also look at the map (\ref{sigma_dual2M}) between fields $\phi$ and $\tilde \phi$ asymptotically:
for $-1 \gg \ka \phi$, we have $\ka \tilde \phi \approx \sqrt3 \log | \phi | \to - \infty$, and for $\phi \to - \infty$ we have $\tilde \phi \approx - \phi \to + \infty$. (So the flow of $\tilde \beta(\tilde\phi)$ should be read in the opposite direction when as function of $\phi$.)

Another interesting example of Type II has been discussed in \cite{Gursoy:2016ggq}, with 
\be
V(\phi) =  \left[\frac{s ( s - d) + d(d-1) v^2}{2\ka^2 \ell^2}\right] \phi^2 - \frac{d ( d-1)}{\ka^2 \ell^2} \cosh( v \ka \phi) .
	\label{PotnewithInstex}
\ee
This potential has rich QCD phenomenology for finite temperature solutions (i.e. black hole geometries); it has AdS asymptotics (\ref{AdSVnearquaphi}) for $\ka\phi \ll 1$, and for $\phi \to \infty$ it goes to 
$V(\phi) \sim -  e^{v \ka \phi}$.

\vspace{3mm}

Finally, there are the flows of Type III, which are ``asymptotically symmetric'' under CSFI, and both $\beta(\phi)$ and $\tilde \beta(\tilde \phi)$ have a UV fixed point and a special-Liouville IR asymptotic.
A first example of this kind is given by the potential (\ref{PotnewithInstex}) with $v = \sqrt2 v_c$, which is precisely the numeric example examined in \cite{Kiritsis:2016kog}, in connection with the dynamical instability of \cite{Gursoy:2016ggq}.
A more restrict example is given by the class of invariant models (\ref{SfDuWphi}).
Here we can give  a remarkable example: the single-field version of the GPPZ flow \cite{Girardello:1999bd} (cf. \cite{Papadimitriou:2004rz,Bianchi:2001kw,Bianchi:2001de}).
This is a relevant deformation of $\cal N = 4$ SYM in the UV, going to a IR fixed point given by $\cal N = 1$ SYM, in $d = 4$. The most general superpotential has in fact  a scalar $m$ which is a singlet of SO(3) and also an SU(3) singlet $\s$ (corresponding to a gaugino condensate), and reads
$W = \frac{3}{4} ( \cosh \frac{2m}{\sqrt3} + \cosh 2\s)$. Here we consider the consistent single-field truncation that sets $\s = 0$, and we call $m = \phi$. (We also use the name `GPPZ flow' for the arbitrary dimension $d$.) 
The superpotential is
\be
\begin{split}
W (\phi) &= \frac{d-1}{2\ka \ell} \left[ 1 + \cosh \Big(\frac{\ka \phi}{\sqrt{d-1}}  \Big) \right]
\\
		&= \frac{d-1}{\ka \ell} \cosh^2 \Big( \frac{\ka \phi}{2\sqrt{d-1}}  \Big).
\end{split}
	\label{GPPZfloW}
\ee
The AdS boundary is at $\phi = 0$ and the special-Liouville singularity at $\phi \to + \infty$, near which $W \sim  e^{v \ka \phi}$
with $v = \sqrt2 v_c$.
The invariance of the GPPZ solution under CSFI can be immediately seen from the fact that we are dealing with (\ref{SfDuWphi}) for $\a = 1/2$, but let us perform an explicit calculation.
The solution in conformal coordinates is 
\be
a(z) = \tan (-z/ \ell) \, , \quad  \sinh \Big( \frac{\ka\phi}{2\sqrt{d-1}} \Big) = \cot (- \tfrac{1}{\ell} z ), 
	\label{SolGPPZaphi}
\ee
with $ -\tfrac{\pi}{2} \ell < z < 0$,
and the beta-function
\be
\beta(\phi) = \frac{2\sqrt{d-1}}{\ka} \tanh \Big( \frac{\ka \phi}{2\sqrt{d-1}}  \Big).
	\label{betaphiGPPZ3}
\ee
Scale factor inversion, with $\tilde z=-z$ and $\c = 1$,
gives 
$\tilde a(\tilde z)= \cot ( \frac{\tilde z}{\ell} )$, with $0 < \tilde z < \tfrac{\pi}{2} \ell$.
Integrating Eq.(\ref{sigma_dual2M}), we get
$
\cot \big( - \frac{\ka \tilde\phi}{2\sqrt{d-1}}  \big) = \cosh \big(\frac{\ka \phi}{2\sqrt{d-1}} \big) ,
$
with $-\infty<\tilde\phi<0$.
Then with Eq.(\ref{WatsnfSFD}) we find the superpotential
\be
\tilde W(\tilde\phi) = \frac{d-1}{\ka \ell } \cosh^2 \Big(-\frac{\ka \tilde\phi}{2\sqrt{d-1}}\Big)
\ee
which has  exactly the same form as (\ref{GPPZfloW}).

\subsection{Renormalized actions and one-point functions} \label{SectRenorActio}

An important part of the holographic renormalization procedure is the calculation of sourced one-point functions, from which higher-point functions can be found by functional differentiation. Here we focus on the scalar one-point function of an operator dual to the scalar field, in (possibly asymptotically) Liouville models.\footnote{In this section we set $\ka^2 = 1$ for simplicity.}

For asymptotically linear-dilaton solutions in the Jordan-frame, described in \S\ref{SectNonConfpBrn}, the one-point functions have been calculated exactly in \cite{Kanitscheider:2008kd}, using the asymptotic expansion (\ref{FGexapn}). Much like in the standard case (where the solution is AdS with a scalar field in the Einstein frame) \cite{Bianchi:2001de,Papadimitriou:2004ap}, 
the one-point functions are related to the ``free'' constants in the asymptotic expansion (\ref{FGexapnd}).
Precisely, $\kappa^{(0)}(x)$ acts as the source of the scalar operator $\scr O_{\Phi}$ dual to the Jordan-frame field $\Phi$, and $\kappa^{(2\s)}(x)$ gives its VEV; the boundary function $\varphi^{(2\s)}(x)$, which is only present for integer $\s$, is related to an anomaly as in \cite{Henningson:1998gx}.
Characteristically, in the linear-dilaton solutions the source enters the one-point function via an \emph{exponential}, viz.
\be
\langle \scr O_{\Phi}(x) \rangle_{\kappa^{(0)}} = {\cal N} \kappa^{(2\s)}(x) \exp \left[\kappa^{(0)}(x) \right],	\label{VEVske}
\ee
 where $\cal N$ is numerical factor.%
 \footnote{%
 This is in contrast with the standard case where the source enters the one-point function ``linearly'', i.e.~as $\langle \scr O \rangle \sim f^{(2n)} + C(f^{(0)})$ for some function $C$, where $f^{(2n)}$ and $f^{(0)}$ play the role of $\kappa^{(2\s)}$ and $\kappa^{(0)}$ in the asymptotic expansion; see e.g.~\cite{Skenderis:2002wp} Eq.(4.9).
 }

The holographic renormalization performed in \cite{Kanitscheider:2008kd,Kanitscheider:2009as} relies on the asymptotically locally-AdS geometry of the Jordan-frame solution, or on a higher-dimensional AdS solution, to which the exponential potential in the Einstein frame is related via generalized dimensional reduction, as discussed in \S\ref{SectNonConfpBrn2}.
These constructions only work for $\a \leq d/2$, cf.~Eq.(\ref{bounaleqd2}), and CSFI relates models where this bound holds to models where it is violated, see (\ref{Intevaagvvc}).
More recently, in \cite{Kiritsis:2014kua}, a formalism was proposed which allows the holographic renormalization in some models without an AdS boundary. This class of models includes asymptotically Liouville models (in the Einstein frame) for the whole range (\ref{vvmaxbonud}), so it can be used for both models  of a CSFI-related pair.

The procedure in \cite{Kiritsis:2014kua} for homogeneous backgrounds 
is very similar to the well-known procedure for asymptotically AdS theories \cite{Bianchi:2001de,Papadimitriou:2004rz}.
If one defines the UV (IR) limit by $e^A$ going to infinity (zero) --- which is consistent with the interpretation of the scale factor as the renormalization energy scale $E$ ---, then in theories (defined by the potential $V(\phi)$) where the superpotential equation (\ref{V1M}) has an \emph{attractor}, in the sense that for any two solutions $W_1(\phi)$ and $W_2(\phi)$ of (\ref{V1M})  
\be
\lim_{\phi \to \phi_\UV} \frac{ W_2(\phi) - W_1(\phi)}{W_2(\phi)} = 0,	\label{attrct}
\ee
one can use the attractor behavior to isolate the counterterms and renormalize the on-shell action.
It is actually not difficult to isolate the attractor behavior. 
Let $W_2(\phi) \equiv W(\phi)$ and $W_1(\phi) \equiv W(\phi) + w(\phi)$ be two solutions of (\ref{V1M}) in the UV limit.
It is not hard to show that  
$w(\phi) = C_w e^{ - d \cal A(\phi) }$,
where $C_w$ is an integration constant and
\be
\cal A(\phi) \equiv - \frac{1}{2(d-1)} \int_{\phi_0}^\phi \! d \varphi \, \frac{W(\varphi)}{\pa_\varphi W(\varphi)} .
\ee
$\phi_0$ is an \emph{arbitrary} boundary condition; a change in $\phi_0$ can be absorbed into $C_w$.
The renormalized action is obtained by subtracting from the on-shell action 
$S_\onsh = 2 \int \! d^dx\, e^{d A(r)} W(\phi)$ a counterterm action $S_{ct} = - 2 \int \! d^dx\, e^{d A(r)} W_{ct}(\phi)$ given by a superpotential $W_{ct}(\phi)$. The attractor behavior then ensures that
\be
S^{\rm{(ren)}} [ E, \phi(E)] = C_R \int \! d^dx \exp \Big[ d \left[ A(E)  - \cal A(\phi(E))  \right] \Big] ,
\ee
where $C_R$ is a scheme-dependent constant because it includes an integration constant from the $W_{ct}$ contribution. 

It is immediate to check, using Eq.(\ref{betaM}), that $S^{\rm{(ren)}}[E , \phi(E)]$ is scale-invariant, i.e.~that 
$$
d S^{\rm{(ren)}} / d\log E = 0 ,
$$
 an indication of the fact that $S^{\rm{(ren)}}$ is valid all along the flow.
For any solution $\{\phi(z),  A(z)\}$ of the field equations, $\cal A(\phi(z))$ and $A(z)$ do coincide in the UV up to an additive constant, which is the reason why the renormalized action is finite.
However, it is important to stress that $\cal A(\phi)$ is a function of $\phi$ only,  independent of the particular radial evolution $\phi(z)$. 
Thus the renormalized action can be written as a function of the energy scale $E$ and of the coupling $\phi$ independently, as 
\be
S^{\rm{(ren)}} [ E, \phi] = C_R \int \! d^dx \exp \Big[ d \left( A(E) - \cal A(\phi) \right) \Big]	\label{Srenzd}
\ee
where it should be understood that $\phi$ varies while keeping $E$ fixed, so we can find \cite{Kiritsis:2014kua} the renormalized one-point function of the operator $\scr O$ sourced by $\phi$, at energy $E$, 
\be
\langle \scr O_\phi \rangle_{E} =  \frac{\delta S^{\rm{(ren)}}}{\delta \phi} = \frac{d C_R}{2(d-1)} \frac{W(\phi)}{\pa_\phi W(\phi)} e^{- d \cal A(\phi)} \, e^{d A(E)} .	 \label{VEVogen}
\ee

In Liouville models, the attractor condition (\ref{attrct}) is satisfied whenever the exponential potential satisfies the bound (\ref{vvmaxbonud}). (See \cite{Kiritsis:2014kua} Sect.7.2.1.) Hence we can use Eq.(\ref{VEVogen}) for both models in a CSFI pair (\ref{VxpVexpSF}).
Inserting the superpotentials explicitly, 
\be
\langle \scr O_\phi \rangle_{E} = \frac{d C_R}{(d-1)v}  \exp \left[ \frac{d(\phi - \phi_0)}{(d-1)v} \right] \, e^{d A(E)}  .	\label{VEVLivaE}
\ee
The dependence of the energy-scale with the scale factor is the usual, viz. $e^{A(E)} = E$ (modulo an arbitrary multiplicative constant), but writing it as a function of $A$ allows us to see clearly, now, the effect of CSFI: taking into account the field transformation (\ref{scfFileLiouvSF}) and scale factor inversion, 
\begin{equation*}
\begin{split}
\frac{d C_R}{(d-1)\tilde v}  &\exp \left[ \frac{d(\tilde \phi - \tilde \phi_0)}{(d-1)\tilde v} \right] \, e^{d \tilde A(\tilde E)} =
	 \frac{d \tilde C_R}{(d-1) \tilde v}  \exp \left[- \frac{d(\phi - \phi_0)}{(d-1)v} \right] \, e^{- d A(E)} .
\end{split}
\end{equation*}
This can be written as
\be
\langle \tilde{\scr O}_{\tilde \phi} \rangle_{\tilde E} = \tfrac{1}{\sqrt{\frac{2 v_c^2}{ v^2} - 1}} \langle \scr O_{-\phi} \rangle_{\frac{1}{E} } .
\ee 
 We see that  one-point functions depend on the fields in a strong-/weak-coupling relation, and are evaluated at reciprocal energy scales.
 
The most usual definition of the one-point function of a scalar operator holds at the UV limit, not at an arbitrary scale of the RG. Instead of (\ref{VEVogen}), the functional derivative is divided by a factor of $\sqrt{-\ga}$, where $\ga$ is the determinant of the induced metric of the  radial slice where the QFT is placed. Since $\sqrt{-\ga} = e^{d A}$, we get 
\be\begin{split}
\langle \scr O_\phi \rangle &= \frac{1}{\sqrt{-\ga}} \frac{\delta S^{(\rm{ren})}}{\delta \phi}
\\	
	& = \frac{d C_R}{(d-1)v}  \exp \left[ \frac{d(\phi - \phi_0)}{(d-1)v} \right] \,   .	\label{VEVLivaE2}
\end{split}\ee
It is interesting to note how the source $\phi$ appears in an exponential, the same structure seen in Eq.(\ref{VEVske}).
Using the freedom to change $\phi_0$ by redefining the unspecified scheme-dependent constant $C_R$, we can formally set $\phi_0 = \phi|_\UV$ and $\phi - \phi_0 =  \cal J$, such that $\cal J = 0$ in the UV. Then 
\be
\langle \scr O_{\cal J} \rangle  = \tfrac{d}{(d-1)v} \cal C_R e^{\frac{d}{(d-1)v} \cal J} ,	\label{VEVJ}
\ee
where now $\cal C_R$ and $\cal J$ play more clearly the part of $\kappa^{(2\s)}$ and $\kappa^{(0)}$. 
In the UV limit, defined as $e^{A} \to \infty$, we can see from Eq.(\ref{aofphexpW}) that $\phi \to - \infty$, hence Eq.(\ref{VEVLivaE2}) shows that $\langle \scr O_\phi \rangle = 0$. Accordingly, $\cal C_R = 0$. 
In the Jordan-frame, the VEV $\langle \scr O_\Phi \rangle$ also vanishes in the exactly linear-dilaton solution (\ref{expPhizz0sol}) where $\kappa^{(2\s)} = 0$.
(Hence in both cases, we find there is no symmetry breaking.)
We note that the constant $C_R$ is undetermined by the method above. In contrast, the renormalization procedure of \cite{Kanitscheider:2008kd,Kanitscheider:2009as} gives an exact one-point function once one finds $\kappa_{(2\s)}$ and $\kappa_{(0)}$ by asymptotically expanding a given AdS linear-dilaton solution. 
Note also that, although it is expected that they are related, the operators $\scr O_\Phi$ and $\scr O_\phi$ are not the same, and the comparison between Eqs.(\ref{VEVLivaE2})-(\ref{VEVJ}) and Eq.(\ref{VEVske}) should be made with care.
In particular, while $\Phi$ is a dimensionless field, the canonical dimension of $\phi$ is $\ka^{-1}$, so the quantization of the corresponding operators may be non-trivially related.  (It would be interesting to explore more deeply the dictionary between holographic renormalization in the Einstein and Jordan frames.) 

We have been considering only homogeneous domain walls, but the asymptotic expansion (\ref{FGexapn}) generally involves functions of the transverse coordinates $x^a$. 
In the Jordan-frame picture, the counter-terms involving transverse derivatives is found by solving algebraically the coefficient-functions in the expansion (\ref{FGexapn}), leading to
\be
\begin{split}
 S^{(ct)} = - \int_{\Sigma_r^h} \!\! d^d x \sqrt{-h} \,
	& e^{\Phi} \Bigg[ 
	2(d - 1) - 4 \a
	\\
&	
	+ \frac{1}{d - 2(\a +1)} \Big[ \scr R^{(d)}(h) + \left(1 + \tfrac{1}{2\a} \right) h^{ab} \pa_a \Phi \pa_b\Phi \Big]
	\Bigg]
\end{split}\ee
see Eq.(6.50) of \cite{Kanitscheider:2008kd}.
The integral is evaluated at a surface $\Sigma_r^h \equiv \{r = \text{constant}\}$, with induced metric $h_{ab}$ and intrinsic Ricci curvature $\scr R^{(d)}$, which is a surface in the radial ADM-like foliation of the spacetime in coordinates $\{r,x^a\}$ where the lapse function is constant.%
\footnote{%
These are related to conformal coordinates by $r = \int^r a(z) dz$.}
In passing to the Einstein frame with (\ref{PhiphiGgtrans}), we find
\be
\begin{split}
S^{(ct)} &=  - 2 \int_{\Sigma_r^\ga} \!\! d^d x \sqrt{- \ga} \,
	\Bigg[ 
	W_{ct}(\phi)
	- U_{ct}(\phi) \Big[  R^{(d)}(\ga) + b_3 \ga^{ab} \pa_a \phi \pa_b\phi \Big]
	\Bigg]
\end{split}\ee
where $\phi = - v_c^2 \Phi / v$, $\ga_{ab} = e^{- v \phi} h_{ab}$ is the induced metric in the corresponding radial surface $\Sigma^\ga_r$ with Ricci curvature $R^{(d)}$, and 
\begin{align}
W_{ct} (\phi) &\equiv B_1 e^{v \phi /2 }
\\
U_{ct} (\phi) &\equiv B_2 e^{- v \phi / 2} \sim 1 / W^{ct}(\phi)
\end{align}
where $B_{1,2}$ and $b_3$ are constants.
The function $W_{ct}$ has the behavior of an exponential superpotential, corresponding to the renormalization of the background (homogeneous) on-shell action (\ref{Srenzd}). The function $U_{ct}(\phi)$ multiplies the terms with transversal derivatives, and has been found (with a slightly different procedure) in \cite{Kiritsis:2014kua}, cf. Eq.(5.3) ibid.
Using the attractor mechanism, one can use these counter-terms to calculate the renormalized action, see Eq.(7.47) of \cite{Kiritsis:2014kua}, 
\be
\begin{split}
S^{\rm{(ren)}} [\phi , \ga_{ab}] &=  \int d^d x \sqrt{- \ga} \Bigg[ C_R \exp \Big( \tfrac{d}{(d-1)v}  \phi  \Big) 
\\
			& +  D_R \exp\Big(  \tfrac{d-2}{(d-1)v} \phi \Big)  \left(  R^{(d)}(\ga) + \tfrac{d_R}{v^2}  \ga^{ab} \pa_a \phi \pa_b \phi \right) \Bigg] 
\end{split}			
			\label{Srenexpsd}
\ee
where $C_R$, $D_R$ and $d_R$ are numerical constants. Eq.(\ref{Srenzd}) corresponds to a zero-derivative limit where the $d$-dimensional curvature $R^{(d)}(\ga)$ vanishes and $\phi$ depends only on the radial coordinate.
As argued in \cite{Kiritsis:2014kua}, the renormalized action  (\ref{Srenexpsd}) is, again, valid at any energy scale along the RG flow,
as it is typical of Hamilton-Jacobi constructions which the formalism in \cite{Kiritsis:2014kua} is an example of.
The renormalized action along the RG flow obtained with Hamilton-Jacobi technology is originally given in \cite{Kanitscheider:2008kd}, see Eq.(6.26). There $S^{(\rm{ren})}$ is written in terms of the highest term of an expansion of the extrinsic curvature in eigenfunctions of the bulk dilatation operator (which is basically the radial derivative $\pa_r$).
The advantage of formula (\ref{Srenexpsd}) for our purposes is that it makes very clear how to apply the CSFI transformations.%
\footnote{%
We thank an anonymous referee for generous remarks on the contents of this Section.}
 CSFI relates different Liouville models with parameters $v$ and $\tilde v$ according to Eq.(\ref{qtqvtvtranf}), so it should act on $S^{\rm{(ren)}}$ in the same way. This is, indeed, the case: the transformation of the fields (\ref{scfFileLiouvSF}) is precisely the correct one to make the functional form of $S^{\rm{(ren)}}$ invariant, only changing $v \leftrightarrow \tilde v$ and $\phi \leftrightarrow \tilde \phi$ because only the combination $\phi / v$ appears in (\ref{Srenexpsd}). 
The fact that CSFI transformations are consistent with the full non-homogeneous action (\ref{Srenexpsd}) is a presage of the fact they can be extended to fluctuations around the isotropic backgrounds in Liouville models --- hence we can indeed expect CSFI to preserve the form of the  $x^a$-dependence of $S^{\rm{(ren)}}$. This extension is what we discuss next.

\section{CSFI for fluctuations and their spectra}	 \label{SectFluctuandSpec}

We now turn to the effect of CSFI on fluctuations around the domain wall geometry. We will be interested in tensor and scalar modes, which are the most relevant for holography.
The linearized metric is%
\footnote{%
The description of fluctuations around domain walls is (basically) the same as in cosmology, see e.g. 
\cite{Mukhanov:1990me}. For domain wall conventions, see e.g. \cite{Kiritsis:2016kog}.}
\be
\begin{split}
& d s^2 = e^{2A} \Big[ [ 1 - (d-2) B ] dz^2 + \left[(1 + B) \eta_{ab} + h_{ab} \right] dx^a dx^b \Big] 
		\label{MetricNwetongau}
\\
& \pa_a h_{ab} = 0 , \quad  h^a{}_a = 0 ,
\qquad
\phi(z, x^a) = \bar \phi(z) + \chi(z, x^a) 		
\end{split}
\ee
Tensor modes $h_{ab}(z,x^a)$ are transverse and traceless, 
and the scalar sector is written in Newtonian gauge. 
We mark the background field with an overbar. 
Since we are working with a single scalar field, there is only one true scalar degree of freedom.
The `Bardeen potential' $B(z, x^a)$, and the perturbation of the scalar field $\chi(z, x^a)$ combine into the gauge-invariant `curvature perturbation' \cite{Maldacena:2002vr}
$\zeta = B - (A' / \bar \phi') \chi$; we use $\zeta$ to describe the scalar d.o.f. 
The linearized field equations are
\begin{align}
h_{ab}'' + (d-1) A' \,  h_{ab}' + \square_d h_{ab} &= 0 
		\label{KGfrohabmai}
\\
\zeta''  +  \left[ (d-1) A'(z)  + \pa_z \log \beta^2(z) \right] \zeta' + \square_d \zeta &= 0 
 		\label{zetaWqsEbta}
\end{align}
where $\beta(\bar \phi)$ is given by Eq.(\ref{betaM}).

\subsection{CSFI and S-duality}	\label{SectSFDSdual}

Eqs.(\ref{KGfrohabmai}) and (\ref{zetaWqsEbta}) can be obtained from an effective quadratic action $I$. Writing $f$ for either the scalar mode $\zeta$ or the tensor modes $h$ (ignoring polarization indexes), 
\bsub\label{ActFlcut}\be
I = \frac{1}{2} \int \! d^d x \!\! \int \! dz \ G(z) \left( f'^2 - \eta^{ab} \pa_a f \pa_b f \right) 
\ee
where
\be
G_h(z) = e^{(d-1) A(z)} ,
\quad
G_\zeta(z) = e^{(d-1) A(z)} \beta^2(z) ,
\ee\esub
and if we pass to phase space with the canonical momentum $\Pi = \delta I / \delta f' = G f'$ we get a Hamiltonian
\be
\begin{split}
\scr H &= \frac{1}{2} \int \! d^d x \, dz \left[ \frac{\Pi^2}{G} + G (\pa f)^2 \right]
\\
	&= \frac{1}{2} \int \! d^d k \, dz \left[ \frac{|\Pi_k|^2}{G} + G k^2 |f_k|^2 \right] .
\end{split}
	\label{HamilFlcu}
\ee
We have used Fourier modes in transverse space, with $|f_k|^2 = f_k f_{-k}$, etc. and $k^2 = \eta_{ab} k^a k^b < 0$ is a timelike vector. 
In \cite{Brustein:1998kq} it was noted that (\ref{HamilFlcu}) is invariant under the transformation 
\be
\begin{split}
G &\mapsto \tilde G = \frac{\la^2}{G}, 
\\
f_k &\mapsto \tilde f_k =  \frac{1}{\la k} \Pi_k, 
\\
 \Pi_k &\mapsto \tilde \Pi_k = - \la k f_k 
\end{split}
	\label{Sdualit}
\ee
where $\la$ is a parameter. This was called \emph{S-duality}.
It is a ``canonical transformation'', exchanging $\Pi$ and $f$; it preserves the Hamilton equations and it swaps the two second-order equations for $f$ and $\Pi$, viz.
\be
\begin{split}
f_k'' + (G'/G) f_k' - k^2 f_k  &= 0 \ ,
\\
\Pi_k'' - (G' / G ) \Pi_k' - k^2 \Pi_k  &= 0 ,
\end{split}
	\label{Eqforffpi}
\ee
the first of which corresponds to Eqs.(\ref{KGfrohabmai})/(\ref{zetaWqsEbta}).

Now, for tensor modes we have $G_h(z) = [ a(z) ]^{d-1}$. Hence $G_h \mapsto \tilde G_h$ is an inversion of the scale factor in conformal coordinates: it is CSFI.
More precisely, \emph{CSFI can be extended to tensor modes as}
\be
\begin{split}
a &\mapsto \tilde a = \frac{\c^2}{ a }
\\
 h_k &\mapsto \tilde h_k =  \frac{1}{\c^{d-1}} k^{-1} \Pi^{(h)}_k 
\\
  \Pi_k &\mapsto \tilde \Pi^{(h)}_k = - \c^{d-1} k \ h_k 
\end{split}
	\label{SdualitTen}
\ee

For scalar modes, the same is not true in general, because inversion of $G_\zeta (z) = [a(z)]^{d-1} \beta^2(z)$ does not correspond to an inversion of $a(z)$. However, in Liouville models, where $\beta$ is a constant, $G_\zeta \mapsto \tilde G_\zeta$ is \emph{again} an inversion of the scale factor: again, it corresponds to CSFI.
Therefore,  \emph{CSFI can be extended to scalar modes in Liouville models as}
\be
\begin{split}
a &\mapsto \tilde a = \frac{\c^2}{ a }
\\
\zeta_k &\mapsto \tilde \zeta_k =  \frac{1}{\beta^2 \c^{d-1}} \ k^{-1} \Pi^{(\zeta)}_k 
\\
\Pi^{(\zeta)}_k &\mapsto \tilde \Pi^{(\zeta)}_k = - \beta^2 \c^{d-1} \  k \ \zeta_k 
\end{split}
	\label{SdualitScal}
\ee
with $\beta = \rm{constant}$.
When a domain wall is only asymptotically Liouville , as in the examples of Sect.\ref{SectSFDasdulflows}, the transformation for the scalar modes is valid asymptotically. 

The name S-duality was given in \cite{Brustein:1998kq} because (\ref{Sdualit}) is a generalization of the usual strong-/weak-coupling duality of string theory.
This is in fact also true for CSFI in the examples we have discussed. 
As a function of the string-frame dilaton $\phi_S$, the Liouville solution (\ref{aofphexpW}) gives (cf.  (\ref{footnoteSframe}))
\[
G \sim a^{d-1} =  e^{- \ka \phi / v}
	=  e^{- 2 v_c \phi_S / v}  .
\]
Hence $G \mapsto 1/G$ is equivalent to $\phi_S \mapsto - \phi_S$, i.e. it is indeed equivalent to an inversion of the string coupling $g_S^2 = e^{\phi_S}$.

\subsection{Schr\"odinger equations}  \label{SectSpectralComput}

It is well known that fluctuations around domain walls can be described by a one-dimensional Schr\"odinger problem with a potential dictated by the background dynamics.
 With a change of variables,
\be
\psi_k (z) \equiv e^{-\int  \cal W(z) dz}  f_k(z) ,
\quad
\cal W(z) \equiv - \tfrac{1}{2} G'(z) / G(z) ,
\label{psiftofca}
\ee
Eq.(\ref{Eqforffpi}) becomes a Schr\"odinger equation with energy $M_k^2 = - k^2 > 0$,
\begin{flalign}
&& H_1 \psi_k &= \left[ - \tfrac{d^2}{dz^2} + \cal V(z) \right] \psi_k = M_k^2 \psi_k , &&		\label{SchroEqQQ}
\\
\text{where} && \cal V(z) &= \cal W^2(z) - \cal W'(z) . &&	\label{SchroPotne}
\end{flalign}
This Hamiltonian is factorizable as 
\be
H_1 = Q^\dagger Q , \quad  Q = \tfrac{d}{dz} + \cal W (z) , \quad Q^\dagger = \tfrac{d}{dz} - \cal W(z) ,
\ee
and has a ``superpartner'' $H_2 = Q Q^\dagger$ whose potential has a flipped sign, 
\be
 H_2 \psi_k = Q Q^\dagger \psi_k =  - \psi_k'' + \left[ \cal W^2(z) + \cal W'(z)\right] \psi_k .
\ee
The special factorization relates the eigenstates and the spectra of $H_1$ and $H_2$. If $\psi_k^{(A)}$ are the wave-functions of $H_A$, 
\begin{align}
\begin{split}
Q^\dagger Q \psi^{(1)}_k = M_k^2 \psi^{(1)}_k
\\
Q Q^\dagger \psi^{(2)}_k = M_k^2 \psi^{(2)}_k
\end{split}
\quad
\begin{split}
\psi^{(2)}_k (z) &= M_k^{-1} Q \psi^{(1)}_k
\\
\psi^{(1)}_k (z) &= M_k^{-1} Q^\dagger \psi^{(2)}_k
\end{split}
		\label{SUSYeinfgpsi}
\end{align}
What we have here is known as `supersymmetric quantum mechanics' (SUSY QM) \cite{Cooper:1994eh}; the function $\cal W(z)$ is (also) called `superpotential'.

%
\begin{figure}[t]
\centering
\includegraphics[scale=0.4]{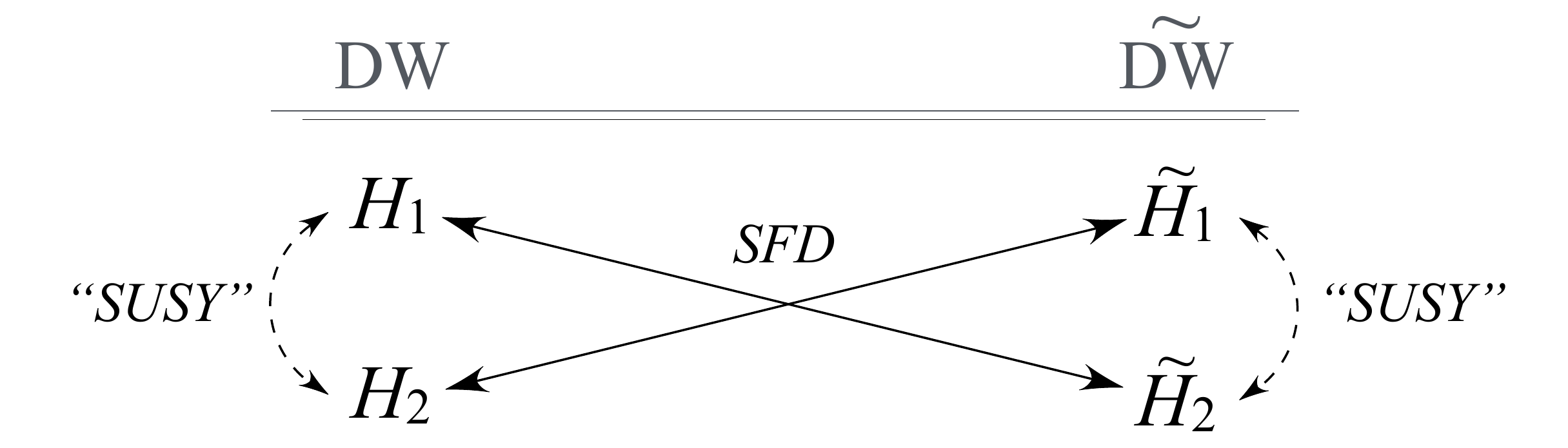}
\caption{Relations between partner Hamiltonians.}
\label{SUSYandSFD}
\end{figure} 
%

We can rephrase the results of \S\ref{SectSFDSdual} in terms of SUSY QM.
The duality of tensor modes written in (\ref{SdualitTen}) here manifests itself in the fact that CSFI is SUSY for the tensor superpotential, which can be read from Eq.(\ref{KGfrohabmai}),
\be
\cal W_T (z) = - \tfrac{1}{2} (d-1) A'(z) . 	\label{SFDcalW}
\ee
Indeed, scale-factor inversion amounts to just flipping the sign of $\cal W_T'(z)$. Therefore CSFI represents a crossed map between the SUSY partners of domain wall solutions related by CSFI, as shown in Fig.\ref{SUSYandSFD}.
If we solve the tensor wave functions around a domain wall $\DW$, (\ref{SUSYeinfgpsi}) gives us automatically the tensor wave functions around its pair $\tDW$ obtained by CSFI. 

The SUSY QM representation is also valid for scalar modes in general, with the SUSY partner of the curvature perturbation $\zeta$ being the Bardeen potential $B$ \cite{Kiritsis:2016kog}. Just as S-duality, the relation between SUSY QM and CSFI, however, only holds for the scalar modes on Liouville backgrounds. 
We can find the scalar superpotential from Eq.(\ref{zetaWqsEbta}),
\be
\cal W_S(z) =  \tfrac{1}{2} (d-1) A'(z) + \pa_z \log | \beta(z) | .	\label{LIovucalWsc}
\ee
In Liouville models $\beta$ is a constant, hence (\ref{LIovucalWsc}) is equal to the superpotential (\ref{SFDcalW}) and CSFI is equivalent to SUSY QM.

The most important fact about the extension of CSFI to fluctuations is that, basically, it exchanges a mode with its derivative. 
In the S-duality context, this corresponds to an exchange between $f_k$ and $\Pi_k$; in the SUSY QM context, it corresponds to the operation $M_k^{-1} Q \psi_k$ leading to a superpartner.  Note that when we express $\psi_k = e^{- \int \cal W dz} f_k$, the corresponding fluctuation/superpartner, given by (\ref{SUSYeinfgpsi}) as
\[
e^{\int \cal W dz} \tilde f_k = \tilde \psi_k
		= M_k^{-1} \left( d/dz + \cal W \right) \psi_k
		= M_k^{-1} e^{- \int \cal W dz} f_k' ,
\]
is equivalent to (\ref{Sdualit}), i.e. to 
$$\tilde f_k = \tfrac{1}{k} \Pi_k = \tfrac{1}{k}  G f'_k$$ 
where by the definition (\ref{psiftofca}) we have $e^{- \int \cal W dz} = G^{1/2}$.
This correspondence between the fluctuation and its derivative has important consequences for the boundary conditions of the pairs of models, as we illustrate concretely below.

\subsubsection{Wave-functions for (asymptotically) Liouville models}	\label{SectExmplLiouvlFlc}

We now apply the CSFI transformations to fluctuations around  asymptotically Liouville (or/and AdS) backgrounds. 
The potentials of the fluctuation equations (\ref{SchroEqQQ}) for both tensor and scalar modes are given asymptotically by
\be
\cal V (z) = \frac{\delta ( \delta - 1)}{(z_* - z)^2} + \rm{O}\left(\frac{1}{|z_*-z|}\right), 
\quad 
\delta \equiv \frac{1}{v^2 - v_c^2} . \label{Vcaloger}
\ee
According to (\ref{qtqvtvtranf}) the parameter $\delta$ transforms as 
\be
\tilde \delta + \delta = 0 ,	\label{delttidelt}
\ee
and in  Table \ref{TabledelvnotAp} we show in each column the ranges of variables related by CSFI.  
The point $z_*$ may be the location of a timelike boundary or of a timelike singularity, depending on whether  $\delta < 0$ or   $\delta > 0$.

\begin{table}
\begin{center}
\begin{tabular}{l*{2}{l}c}
Timelike Boundary			\qquad	&	\qquad	 Timelike Singularity  
\\
\hline
&
\\
$v^2 \in (0 , v_c^2)$		\qquad		& \qquad		$v^2 \in (v_c^2 , 2 v_c^2)$   
\\
$\delta \in (-\infty , - 1/v_c^2)$	\qquad	& \qquad		$\delta \in (1/v_c^2 , + \infty)$   
\\
$z < z_*$	\qquad					&	\qquad	$z > z_*$ 
\\
&
\end{tabular}
\end{center}
  \caption{Ranges of $\delta$, $v$ and $z$ in (asymptotically) Liouville models.}
\label{TabledelvnotAp}
\end{table}

The asymptotic solution near $z_*$  is
\be
\begin{split}
\psi^{(1)}_k (z) &= C_+(k) (z - z_*)^{ \delta} \big[ 1 + b_+(k) (z - z_*)^2  
									+ \rm{O}(z-z_*)^3 \big] 
\\
		& + C_-(k)   (z - z_*)^{1- \delta} \left[ 1 + \rm{O}(z-z_*)^2 \right]
\end{split}
	\label{Expsndonpsidd}
\ee
and its pair can be obtained by applying the $Q$ operator,
\be
\begin{split}
\tilde \psi^{(2)}_k(\tilde z) &= - \frac{(1+2\tilde\delta)C_-(k)}{M_k} (\tilde z_* - \tilde z)^{\tilde \delta}  
\\
	&\quad + \frac{2 b_+ (k) C_+ (k)}{M_k}   ( \tilde z_* - \tilde z)^{1- \tilde \delta} + \cdots
\end{split}
	\label{tildepsi2expsnz}
\ee
where $\tilde z_* - \tilde z = z - z_*$.
Of course, (\ref{tildepsi2expsnz}) has the same form as (\ref{Expsndonpsidd}), but with the integration constants related as
\be
\tilde C_+ = - 2 \nu C_- / M_k  , \qquad \tilde C_- = 2 b_+ \, C_+  / M_k .
		\label{IRUVconstFIXED}
\ee
This  shows that, by fixing boundary conditions at $z_*$ in $\psi^{(1)}_k$, the CSFI  image of the boundary condition at $\tilde z_*$ in $\tilde \psi^{(2)}_k$ is univocally determined.
Thus, even when CSFI holds only asymptotically, it relates Dirichlet and Neumann (or mixed) boundary conditions of the paired wave functions.
Going back to the original perturbation with Eq.(\ref{psiftofca}), we find 
\be
f_k(z) \approx \frac{L^{\delta +1}}{|1-v_c^2/v^2|^{\delta +1}} \Big[
	C_+(k) +  C_-(k) \, | z - z_*|^{-2\delta + 1} \Big] .
		\label{metrflucfp}
\ee
In the near-boundary limit we have $\delta < 0$.

In special cases, the fluctuations can be solved exactly. Then CSFI can be used as a ``solution generating technique'' for tensor fluctuations in a similar way as it is for background solutions. We give an illustration of this fact in App.\ref{AppFlowexamplC}, by finding the exact solution of fluctuations around a complicated domain wall.

\subsection{Spectra of bound states} \label{SectNomaWavFunc}

Bound states of the $d$-dimensional QFT occur when $\psi_k$ is normalizable both in the UV and in the IR, i.e. when
\be
\int_{z_\IR}^{z_\UV} \! \! dz \  |\psi_k (z)|^2 < \infty ,
				\label{intmustbefinMa}
\ee
ensuring that the kinetic term of the effective $d$-dimensional action for $f(z,x)$ is finite.

We are, as usual, interested in domain walls which are asymptotically Liouville or AdS.
The first thing we prove is that the quantum-mechanical SUSY  is broken for the potential (\ref{SFDcalW}). 
This means that neither the solution of  $Q\psi_0^{(1)} = 0$ nor the solution of $Q^{\dagger}\psi_0^{(2)}=0$ are square-integrable, which is easily verifiable, since the solutions are
\be
\begin{split}
\psi_0^{(1)}(z) &= C_\UV \exp \left[ \frac{(d-1)}{2} A(z) \right], 
\\
 \psi_0^{(2)}(z) &= C_\IR \exp \left[ -\frac{(d-1)}{2} A(z) \right], 
\end{split}
\ee
and $a(z)$ is given by (\ref{SolforazExp}) and/or (\ref{SolforazExp2}) in the UV and IR asymptotics. Hence $Q$ and $Q^\dagger$ do not change the ``energy levels'', and the partner models have a completely degenerated spectrum \cite{Cooper:1994eh}
\begin{align}
\begin{split}
H_1 \psi^{(1)}_n &= M^2_{(1)n} \psi^{(1)}_n
\\
\tilde H_2 \tilde \psi^{(2)}_n &= \tilde M^2_{(2)n} \tilde \psi^{(2)}_n
\end{split}
\quad 
 \begin{split}
 \tilde M_{(2)n}^2 &=  M_{(1)n}^2 = M_n^2
\\
 M_0^2 &\neq 0  , 
\end{split} 
\label{SUSYWMmpasbr}
\end{align}
with $n = 0, 1, 2, 3, \cdots$
Therefore, given any domain wall with a tensor mass spectrum $\{ M_n \}$, \emph{CSFI yields a different domain wall which has the same spectrum}.

\subsubsection{Stability}	\label{SectSatabe}

Fluctuations are stable as long as $M_k^2 \geq 0$. Starting from  
$\int_{z_\IR}^{z_\UV} dz \ [ Q \psi_k  ]^* [ Q \psi_k] \geq 0$, the special factorization of the SUSY Hamiltonians implies that
$\psi_k^*(z) \ Q \psi_k(z) ]_{z_\IR}^{z_\UV} + M_k^2 \int_{z_\IR}^{z_\UV} dz \ | \psi_k(z)|^2 \geq 0$, hence a sufficient condition for $M_k^2 \geq 0$ is that the boundary term vanishes,
\be
\psi_k^*(z) \, Q \, \psi_k(z) \Big]_{z_\IR}^{z_\UV} = 0 . 	\label{stabilcondbs}
\ee 
Thus imposing Dirichlet or Neumann(-like) boundary conditions,
\begin{flalign}
\text{(Dirichlet)} && \psi_k(z_\IR) = \psi_k(z_\UV) = 0  &&	\label{DirchlCondpsi}
\\
\text{(``Neumann'')} && Q\psi_k(z_\IR) = Q\psi_k(z_\UV) = 0  &&
\end{flalign}
is sufficient to ensure stability of the fluctuations. (There are, of course, other ``mixed'' choices.)
Note that the form of Eq.(\ref{stabilcondbs}) is such that, taking into account (\ref{SUSYeinfgpsi}), if $\DW$ is stable then $\tDW$ will be as well, i.e. CSFI preserves stability.

It is not guaranteed that (\ref{stabilcondbs}) can indeed be imposed. Analyzing the asymptotic solutions of (\ref{SchroEqQQ}), it was shown in \cite{Kiritsis:2016kog} that stability is ensured for domain walls which have an AdS UV boundary within the Breitenlohner-Freedman unitarity bound and a regular (i.e. non-singular) IR, or a singular Liouville IR satisfying the condition
\be
\delta \geq 3/2
\quad
\text{hence}
\quad
v_c^2 < v^2 \leq v_c^2 + \tfrac{2}{3} ,	\label{welldefintev}
\ee
cf. Eq.(\ref{Vcaloger}).
This condition ensures that the Hamiltonian with the Schr\"odinger potential (\ref{Vcaloger}) is self-adjoint.
When (\ref{welldefintev}) holds, we are forced to make $C_- = 0$ in (\ref{Expsndonpsidd}), otherwise the wave function is not square-integrable.%
\footnote{%
For $0 < v^2 < v_c^2$ (thus for $\delta < 0$), the fluctuations are not integrable because then the singularity lies at $z \to \infty$, hence $\cal V(z)$ is not bounded and the spectrum is a continuum.}
Moreover, when (\ref{welldefintev}) holds,  normalizability \emph{forces} the Dirichlet boundary condition $\psi(z_\IR) = 0$,  leaving only one integration constant to be fixed at $z_\UV$. We thus have \emph{a well-defined spectrum imposed by the condition of normalizability of the wave function.}

When (\ref{welldefintev}) does \emph{not} hold, i.e. when $\delta  \in ( 0 , 3/2)$, normalizability does not forcefully imply an IR Dirichlet condition.
(Both terms in (\ref{Expsndonpsidd}) are normalizable.)
Then the Hamiltonian of the Schr\"odinger potential (\ref{Vcaloger}) is not self-adjoint, but there are techniques that allow the construction of families of self-adjoint extensions, described in terms of a continuous parameter
\cite{gitman2012self,Gitman:2009era,CamaradaSilva:2018tpk}.
It can be shown \cite{Bouaziz:2014wxa} that there exists only one negative eigenvalue $M^2$ as long as $\delta (\delta -1) \geq - \tfrac{1}{4}$, and this mode can always be isolated and excluded, thus ensuring the stability of the spectrum.

Now, recall that for potentials with well-defined CSFI pairs, the parameter $v$ is bounded from above by the special Liouville value $0 < v^2 \leq 2 v_c^2$. How does this fit with the interval (\ref{welldefintev})? The answer depends on the dimension $d$. The most important case for holography is $d = 4$, and in this case, the two intervals \emph{coincide}.
This is a limiting case:
for $d \geq 5$,  every Liouville singularity with $v^2 \leq 2 v_c^2$ also satisfies (\ref{welldefintev}), but for $d = 3$ (or 2) the special Liouville singularity does \emph{not} satisfy (\ref{welldefintev}).

\subsubsection{Example: GPPZ flow in $d =3$}	\label{SectGPPZd3}

Consider the GPPZ flow (\ref{GPPZfloW}) in 3+1 dimensions.
This is a useful example for two reasons.
First, it has the peculiar property of being symmetrical under CSFI.
Second, since $d = 3$, it fits the case mentioned above, in which the wave-functions are automatically normalizable in the IR, so one needs an extra criterion for fixing the IR boundary conditions.
We will show how  \emph{the invariance under CSFI can be used to fix the IR behavior of the wave-function in terms of its UV properties.}

The background geometry, given by Eq.(\ref{SolGPPZaphi}), shows an AdS boundary at $z_\UV = - \tfrac{\pi}{2} \ell$ and a Liouville singularity at $z_\IR = 0$.
Given $a(z)$, we find the SUSY QM superpotential (\ref{SFDcalW}), and the potential (\ref{SchroPotne}) for tensor modes:%
\footnote{%
We describe tensor modes, but a similar discussion goes for the scalar ones. Note that the exact invariance under CSFI extends to the scalar modes even if this is not a Liouville model.}
\be
\cal W_T(z) = \frac{2 / \ell }{\sin(-2z/\ell)};
\qquad
\cal V_T(z) = \frac{2 / \ell^2}{ \cos^2 (-z / \ell )}.		\label{calWV1}
\ee
The Schr\"odinger equation (\ref{SchroEqQQ}) has the exact solution
\be
\begin{split}
\psi(z) &= C_+ \left[ \sqrt{M^2}\sin (-\sqrt{M^2}z ) + \tfrac{1}{\ell} \tan ( \tfrac{1}{\ell} z  ) \cos (\sqrt{M^2}z ) \right]
\\
	&+ C_- \left[\sqrt{M^2}\cos (-\sqrt{M^2}z ) - \tfrac{1}{\ell} \tan (\tfrac{1}{\ell} z  ) \sin (\sqrt{M^2} z ) \right]
\end{split}	
	\label{ggpz3genpsi}
\ee
where $C_\pm$ are arbitrary constants to be fixed by the boundary conditions. As it should be expected from our previous discussion, $\psi(z)$ is regular at the IR, with $\psi(z_\IR) =  C_- \sqrt{M^2}$.
Thus normalizability at the singularity does not require neither $C_+$ nor $C_-$ to be zero. 

We then turn to the UV limit, where we must impose Dirichlet conditions such that $\psi(z_\UV) = 0$ (we are after bound states). This fixes the mass spectrum to be $M_k^2 = k^2 / \ell^2$, with $k = 2,3,4,5, \cdots$. The eigenfunctions split into two classes, depending on wether $k$ is even or odd. For odd $k = 2n +3$ we must take $C_- = 0$ and the normalized functions are
\be
\begin{split}
\psi^+_n(z) &= \tfrac{1}{\sqrt{\pi \ell (n+1)(n+2)}} \Big[ (2n+3) \sin\left( \tfrac{2n+3}{\ell} z \right) 
	 - \cos \left( \tfrac{2n+3}{\ell} z \right) \tan\left( \tfrac{1}{\ell} z  \right) \Big]
\\
 M^2_n &= \frac{(2n+3)^2}{\ell^2}\ ; \qquad  n=0,1,2, 3, \cdots 
\end{split}
	\label{psinGPPZ3}
\ee
For even $k = 2n$, we have to take $C_+ = 0$ instead, and the normalized solutions are
\be
\begin{split}
& \psi^-_n(z) = \tfrac{2}{\sqrt{\pi \ell (4n^2-1)}} \Big[ 2n \cos \left( \tfrac{2 n}{\ell} z \right) + \tan \left(\tfrac{1}{\ell} z  \right) \sin \left( \tfrac{2n}{\ell} z \right) \Big]
,
\\
& M^2_n = \frac{4n^2}{\ell^2}
\end{split}	
\label{Gppsevenpsin}
\ee
with $n = 1,2,3, \dots$
Thus we have two sets of eigenfunctions which are regular in the singularity and give discrete spectra in the UV.
The functions $\psi^+$ vanish at $z_\IR = 0$, which is usually the boundary condition imposed by regularity at the singularity. The functions $\psi^-$ do not vanish at $z_\IR$ but they are finite and hence square-integrable, so they cannot be discarded.  

Now we use CSFI. The SUSY QM superpotential and potential are 
\be
\tilde {\cal W}_T (\tilde z) = \frac{2 / \ell^2}{ \sin(2\tilde z / \ell)} ,
\qquad
\tilde {\mathcal V}_T(\tilde z)=\frac{2/\ell^2}{\sin^2(\tilde z / \ell)} 
			\label{calWV2}
\ee
with $0 < \tilde z < \tfrac{\pi}{2} \ell$, and  $\tilde z_\IR = \tfrac{\pi}{2} \ell$ and $\tilde z_\UV = 0$.
Note that the SUSY partners (\ref{calWV1}) and (\ref{calWV2}) are actually identical, there is only a translation by $\tfrac{\pi}{2} \ell$. This is a consequence of the symmetry under CSFI. Using (\ref{SUSYeinfgpsi}), i.e. applying the operator $M^{-1}_n Q$ on $\psi^+_n(z)$,  we find
\begin{align*}
\tilde\psi^+_n (\tilde z) &= \frac{\ell}{2n+3} \left[ \partial_z + \frac{2/\ell}{\sin(-2z/\ell)}\right] \psi^+_n(z) \Bigg|_{z=-\tilde z}
\\
	&= \tfrac{1}{\sqrt{\pi \ell (n+1)(n+2)}} \Big[ (2n+3) \cos\left( \tfrac{2n+3}{\ell} \tilde z \right) 
	- \sin\left(\tfrac{2n+3}{\ell}  \tilde z \right) \cot \left( \tilde z / \ell \right) \Big]
\end{align*}
with $\tilde M^2_n = (2n+3)^2 / \ell^2$ and $n=0,1,2,3 ,\cdots$. These are the same as the original functions (\ref{psinGPPZ3}), which is also consistent with the invariance of the model.

But for the even functions (\ref{Gppsevenpsin}),  the image solutions are
\begin{align*}
\tilde\psi^-_n(\tilde z) &= \frac{\ell}{2n} \left[ \partial_z + \frac{2}{\ell \sin(-2z/\ell)} \right] \psi^-_n(z) \Bigg|_{z=-\tilde z}
\\
	&= \tfrac{2}{\sqrt{\pi \ell (4n^2 - 1)}} \left[ \cos \left(\tfrac{2n}{\ell} \tilde z \right) \cot \left(\tfrac{1}{\ell} \tilde z\right) + 2 n \sin \left( \tfrac{2n}{\ell} \tilde z  \right) \right].
\end{align*}
which are \emph{not} the same as (\ref{Gppsevenpsin}), an inconsistency with invariance.
The point, however, is that these solutions should be discarded. Indeed, they \emph{diverge} at $\tilde z_\UV$, so they do not represent bound states. But to force $\tilde \psi^- = 0$, we must take $\psi^- = 0$, and hence choose $C_- = 0$ in (\ref{ggpz3genpsi}), leaving only the functions (\ref{psinGPPZ3}) as eigenfunctions.
Thus in the end, consistency with the symmetry of the model under CSFI has discarded half of the spectrum allowed by normalizability, by ultimately fixing a specific IR Dirichlet condition.

\section{Conclusion}		\label{SectConclusion}

We conclude this paper with a discussion and some speculations about the nature and the possible uses of conformal scale factor inversion correspondence.

\begin{description}[style=unboxed,leftmargin=0cm]

\item[CSFI as a duality]

The idea of a scale factor inversion map is very similar in nature to the `scale factor duality' (SFD) transformations used in cosmology. The paradigmatic example of scale factor duality is the one found by Veneziano \cite{Veneziano:1991ek}, but other maps between cosmological solutions have also been called by the same apellation, e.g. \cite{Chimento:2003qy,dS:2015fwa,dS:2016kdc,Gurses:2020ndy}.
The analogous of our map for cosmological spacetimes, in particular, has been called a scale factor duality in \cite{dS:2015fwa,dS:2016kdc}. 
In the context of string theory and holography, however, the word `duality' should be used with some care to avoid misinterpretation.

The distinguishing feature of Veneziano's SFD is to be a symmetry of the string effective action (with a constant potential), hence a kind of extension of T-duality for time-dependent (cosmological) backgrounds. 
As discussed in Sect.\ref{SectSFD}, CSFI is not a symmetry of the action (\ref{ActSGVph}), so it cannot be interpreted as a duality in this sense.  
In particular, although it is based on a scale factor inversion, it is not connected with a T-duality transformation, at least not in any evident way. It would certainly be interesting to investigate whether such a connection can be established (this would perhaps require the inclusion of a two-form field).

The strict interpretation of CSFI is as a map between solutions of different theories --- each with a scalar potential. The map has a $Z_2$ structure: applied twice, it gives again the original domain wall solution, $\DW \mapsto \tDW \mapsto \DW$.
For Liouville models, the map has the additional property of being a map in parameter space: schematically, a Liouville model $\rm{LV}_v$ with a parameter $v$ in the exponent is mapped to
$\rm{LV}_v \mapsto \widetilde{\rm{LV}}_{\tilde v} \mapsto \rm{LV}_v$.

Extended to fluctuations around isotropic domain walls, CSFI gives  a canonical transformation swaping a fluctuation mode and its conjugate momentum.
The extension is (always) valid for tensor modes, and is valid also for scalar modes (only) in Liouville models.
This map between fluctuations \emph{can} be interpreted in terms of the S-duality found in cosmological spacetimes by Brustein, Gasperini and Veneziano \cite{Brustein:1998kq}.

The peculiar behavior of Liouville models may be an indication that, perhaps, CSFI could be related more precisely to a kind of ``generalized'' conformal symmetry --- we note that the defining property of CSFI is precisely that the scale factor is a conformal factor in the Einstein frame. 
Understanding this point could shed an interesting light on the nature of invariant models, such as the GPPZ flow, and on the connection between the two geometric constructions discussed in \S\ref{SectNonConfpBrn2} between the complementary classes of dimensional reduction leading to Liouville potentials.

\item[IR boundary conditions]

Fluctuation modes whose wave-functions vanish at the UV boundary correspond to composite particles in the $d$-dimensional QFT, and the exact form of the mass spectrum crucially depends on the boundary conditions to be imposed  at the IR.  Domain walls with a singularity in practice introduce the IR cutoff \emph{if} the normalizability condition (\ref{intmustbefinMa}) implies a natural Dirichlet boundary condition at the singularity.  For Liouville singularities, this is the case when the parameters obey the bound (\ref{welldefintev}). Then the natural IR condition is $\psi( z_\IR) = 0$, or $C_- = 0$ in Eq.(\ref{Expsndonpsidd}), because this is the only way to make $\psi(z)$ square-integrable. Singularities with such a behavior have been called `spectrally computable' in \cite{Kiritsis:2016kog},  and (\ref{welldefintev}) has been called a  `computability bound'. When the computability bound is \emph{not} satisfied, IR normalizability becomes trivial, ceasing to be a useful criterion. Then the spectrum is unspecified unless some different criterion fixes the IR boundary condition, as has been argued by some authors \cite{Kiritsis:2016kog, Gursoy:2007er,Kiritsis:2006ua}.

CSFI transforms a UV boundary condition into an IR one: according to (\ref{IRUVconstFIXED}), a bound state of $\tDW$, i.e. $\tilde \psi(\tilde z_\UV) = 0$, implies the Dirichlet  condition $\psi(z_\IR) = 0$ in the dual  $\DW$.  This could be an interesting complementary  condition to be used when spectral computability fails, but the situation is delicate. We have shown that for $d = 4$ the computability bound is identical with the restriction (\ref{bongvvc}) that selects the Liouville models that possess good (NEC-preserving)  pairs under CSFI. For $d \geq 5$, the computability bound is less restrictive than (\ref{bongvvc}) hence all NEC-preserving models are already computable. Conversely, if we take a domain wall which is not spectrally computable, we \emph{cannot} use CSFI to impose a boundary condition because the dual domain wall will violate the NEC.

In some very specific cases, however, CSFI can be used to provide a IR criterion, by mapping the unspecified IR limit to the UV of a different solution. 
The simplest example of such would be the map of a special-Liouville singularity to an AdS boundary. 
Consider then a special-Liouville IR limit with $\delta \in (0, 3/2)$; square-integrability of $\psi^{(1)}_k$ in Eq.(\ref{Expsndonpsidd}) does \emph{not} impose $C_- = 0$. On the other hand, the image of  $\psi^{(1)}_k$ under CSFI, i.e. $\tilde \psi^{(2)}_k$ in (\ref{tildepsi2expsnz}),  should also be square-integrable. But for $C_- \neq 0$ this only happens if  $\delta=-\tilde{\delta}<1/2$. Hence the consistency of CSFI is more restrictive than the ``normalizability condition'', as the former \emph{does} impose $C_- = 0$ whenever $\delta$ lies not in (\ref{welldefintev}) but in the (larger) interval $\delta>1/2$.
Now, as discussed at the end of \S\ref{SectSatabe},
when $d =3$ we have $\delta > 1$ and when $d = 2$ we have $\delta > 1/2$, so in these dimensions this use of CSFI always fixes univocally the IR boundary condition.
This is, admittedly, an idiosyncratic example, but fixing the computability problem could have some unexpected applications;
in \cite{Scursulim:2020ats} it was shown to be related to a well-known problem in conformal quantum mechanics.

Of course, the operation of fixing IR boundary conditions to be the image of UV boundary conditions becomes truly interesting in models which are invariant under CSFI.
The requirement of symmetry of the fluctuation modes fixes the spectrum uniquely  (and even when the computability bound is violated), 
as shown by the explicit example of the GPPZ flow in $d = 3$ given in \S\ref{SectGPPZd3}.

\item[Holographic cosmology]

The correspondence, under an analytic continuation, between isotropic domain walls and Friedmann-Robertson-Walker spacetimes  \cite{Skenderis:2006jq,Skenderis:2006rr,Skenderis:2006fb} has been used in the construction of a holographic formulation of cosmology \cite{McFadden:2009fg}.
One of the most interesting applications of the results of the present paper is the use of CSFI in this context. 
A first point to be noted here is that the CSFI  transformations hold for any spatial curvature of FRW geometries. In $d = 3$,  CSFI has the interesting property of mapping AdS space to a (special) Liouville model where the energy density of the scalar field scales like radiation, $\frac{1}{2} \dot \phi^2 + V(\phi) \sim a^{-4}$. This could be interpreted as a map between inflation and a radiation-dominated universe \cite{dS:2016kdc}, and it would be very interesting to investigate in detail this connection in the context of holographic cosmology. The aforementioned non-trivial use of CSFI in fixing the boundary conditions of fluctuations in $d = 3$ could play an important role in a cosmological framework.

\end{description}


\vspace{7mm}

\begin{center}
\textbf{Acknowledgements}
\end{center}


\noindent
The authors would like to kindly thank an anonymous referee for illuminating comments and suggestions made on a previous version of this paper.

\vspace{3mm}

\noindent
The work of G.S.~is partially supported by the Bulgarian NSF grant KP-06-H38/11.

\appendix

\section{Causal structure of CSFI pairs}	\label{AppCausalStruc}

\begin{figure}[t] 
\centering
\includegraphics[scale=0.45]{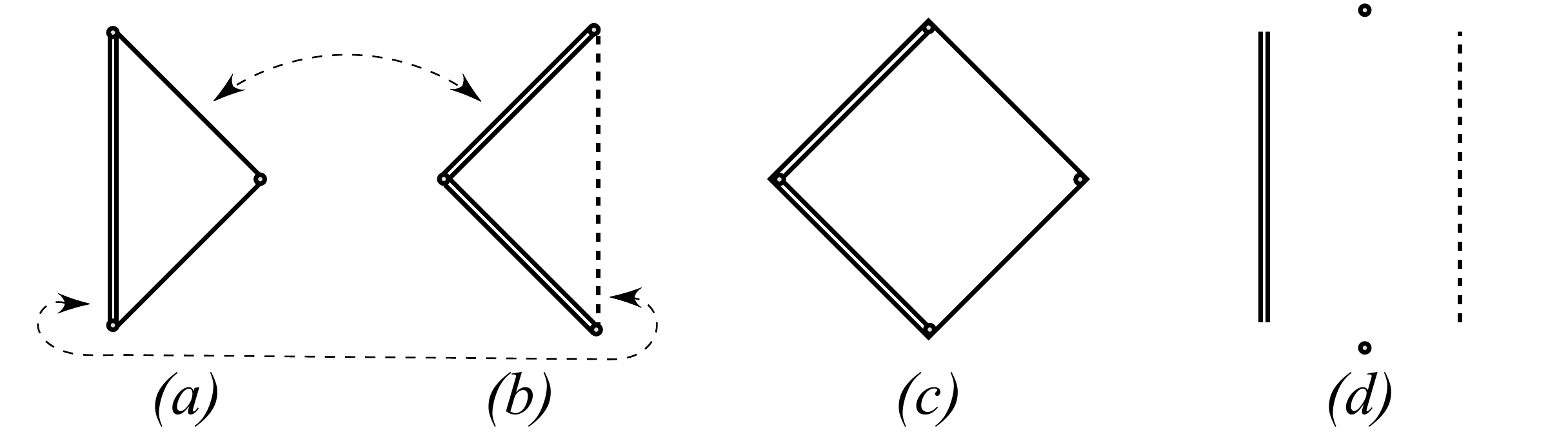}
\caption{Penrose diagrams. Double lines are singularities; dashed lines are timelike boundaries. The arrows indicate the CSFI map.
\textit{(a)} Liouville models with $v > v_c$;
 \textit{(b)} Liouville models with $v < v_c$; 
 \textit{(c)} Liouville models with $v = v_c$; 
  \textit{(d)} Models with an AdS boundary and a Liouville singularity with $v > v_c$. 
}
\label{Penroses}
\end{figure}

Since it is defined in conformal coordinates, CSFI preserves the causal nature of a surface.
Thus, if we generically call $\{a = 0\}$ a ``singularity'' and $\{a = \infty\}$ a ``boundary'', it is not hard to see that CSFI maps 
\begin{center}
null boundaries $\longleftrightarrow$  null singularities

timelike boundaries $\longleftrightarrow$  timelike singularities
\end{center}
This is shown in Fig.\ref{Penroses}.

%
%

The best examples are the Liouville models.
A Liouville model with $v > v_c$ has the singularity at a finite distance which we can set to $z_0 = 0$, so it is timelike, while the boundary lies at $z = - \infty$ so it is null. The Penrose diagram is in Fig.\ref{Penroses}\text{(a)}. On the other hand, its image under CSFI, with $v < v_c$, will have null singularity and a time-like boundary as shown in Fig.\ref{Penroses}\text{(b)}. 
The critical model with $v = v_c$ has boundary at $z = \infty$ and singularity at $z = - \infty$, so both are null and the diagram is that of Fig.\ref{Penroses}\text{(c)}.

The Penrose diagram of an invariant model must be shape-invariant under CSFI. This is illustrated by the Liouville model with $v = v_c$. Meanwhile, an invariant model with an AdS boundary  (e.g. the GPPZ geometry) has a diagram like in Fig.\ref{Penroses}\text{(d)} (which is also invariant because boundary and singularity are both timelike). 

On the other hand, for a model which is not invariant, the shape of the diagram is not preserved.
Fig.\ref{Penroses}\text{(d)} is also the diagram of, say, a domain wall whose potential has two different Liouville asymptotics: with $v > v_c$ as $a \to 0$ and $v < v_c$ as $a \to \infty$. If these limits are reversed, we have a Fig.\ref{Penroses}\text{(c)} diagram, etc.

Note that a discrete spectrum occurs in models where the interval $[z_\IR , z_\UV]$ is finite (then the Schr\"odinger problem becomes a finite box). The casual structure of such domain walls is that of Fig.\ref{Penroses}\textit{(d)} and, as we have shown, this structure is preserved by CSFI. Continuous spectra, on the other hand, happen when there is an infinite, or semi-infinite range of the conformal coordinate $z$, such as in in diagrams \textit{(a)}, \textit{(b)} and \textit{(c)}. Again, this structure is preserved by CSFI.

\section{RG flow from AdS to Critical Liouville, and its image}	\label{AppFlowexamplC}

Here we present another example of background for which the tensor fluctuations can be solved exactly.
The interesting thing about this model is that the image background under CSFI is quite complicated, but the map gives us the corresponding fluctuations nevertheless.

We start with a model given by
\be
\begin{split}
W(\phi) =\tfrac{d-1}{\ka \ell} \cosh \Big(\tfrac{\ka\phi}{\sqrt{2(d-1)}} \Big) \ ,
\qquad
\beta(\phi) = \tfrac{\sqrt{2(d-1)}}{\ka} \tanh \Big( \tfrac{\ka\phi}{\sqrt{2(d-1)}} \Big) .
\end{split}
\ee
There is a UV AdS boundary with radius $\ell$, and with $s = 1$, while for  $\ka\phi\gg1$ we have $W\sim e^{\ka\phi/\sqrt{2(d-1)}}$, thus a \emph{critical} Liouville singularity.
The beta-function coincides with (\ref{betaphiGPPZ3}) near the UV fixed point. (Actually, the two functions coincide after a rescaling of $\phi$.) This illustrates how the same flux may correspond to very different bulk geometries.
Solving the domain wall profile, we find
\be
a(z) = \frac{1}{\sinh(z/\ell)}  , \quad  \ka \phi(z) = \sqrt{2(d-1)}\,z / \ell , 
	\label{aandphicritad}
\ee
with $0 < z < \infty$.
The AdS boundary is at  $z_\UV = 0$, and near the critical Liouville singularity at $z_\IR = \infty$ we have $a(z)\sim e^{-z/ \ell}$; cf. Eq.(\ref{SolforazExp2}).

From (\ref{aandphicritad}) we find the SUSY QM potential and superpotential of the tensor fluctuations to be
\be
\begin{split}
\cal W_T (z) = \frac{d-1}{2\ell} \coth\left(z / \ell \right), 
\qquad
\cal V_T (z) = \frac{(d-1)^2}{4\ell^2} + \frac{d^2-1}{4\ell^2}\frac{1}{\sinh^2(z/\ell)} .
\end{split}\ee
The UV-square-integrable solution of the Schr\"odinger equation, satisfying the Dirichlet boundary condition  $\psi(z_\UV) = 0$ at $z_\UV = 0$, is
\be
\begin{split}
 \psi(z) &= C \left[ \sinh \left(\tfrac{z}{\ell} \right) \right]^{\frac{d+1}{2}} 
	{}_2F_1\left[ \tfrac{d+1}{4} - \tfrac{i g \ell}{2}  , \tfrac{d+1}{4}+\tfrac{i g \ell}{2}   ;  \tfrac{d+2}{2} ;  -\sinh^2 \left( \tfrac{z}{ \ell} \right)\right]
\\
g &=\sqrt{M^2- \tfrac{(d-1)^2}{4l^2}}.
\end{split}
\ee
Near the boundary,  $\psi(z)\approx C \left( z / \ell \right)^{(d+1)/2}$. Near the singularity,
\be
\begin{split}
 \psi(z) &\approx C |\lambda_g| \cos \Big[ g (z-\ell \log2) + \tfrac{1}{2} \delta_g \Big],
\\
 |\lambda_g|  &= \frac{2\Gamma \left(ig \ell \right) \Gamma \left(\tfrac{d+2}{2}\right) e^{-i \delta_g / 2} }{ \Gamma \big(\tfrac{d+1}{4} + \tfrac{ i g\ell}{2}  \big) \Gamma \big(\frac{d+3}{4} - \tfrac{i g\ell}{2} \big)},
\end{split}
	\label{phshsit1}
\ee
where $\delta_g$ is the phase-shift of the wave-function.

\bigskip

\begin{figure}    
    \begin{center} 
        \includegraphics[scale=0.5]{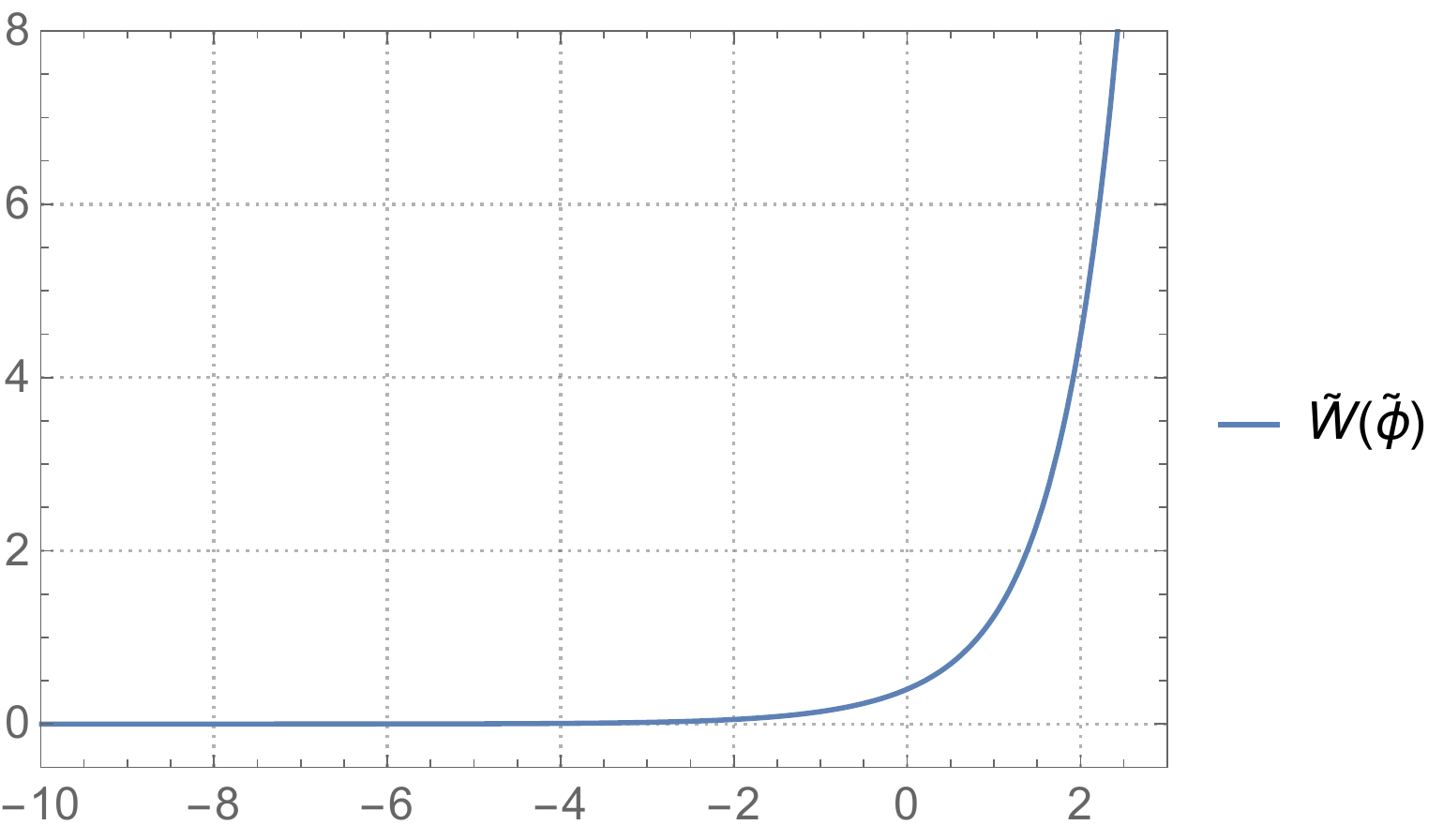}
    \caption{Superpotential (\ref{dualWapp}).}        \label{PlotForApp}
    \end{center}
\end{figure}

Now, with CSFI we are going to construct a new domain wall, with a special-Liouville singularity (image of the AdS boundary) and an asymptotic critical-Liouville boundary (image of the critical Liouville singularity).
Using $\c = 1$ and $\tilde z = - z$, we have the  scale-factor
\be
\tilde a(\tilde z) = \sinh \left(- \tilde z / \ell \right), \qquad - \infty < \tilde z < 0 .
	\label{tilazcsra}
\ee
The scalar field can be expressed implicitly as
\be
\begin{split}
&\ka \tilde\phi = \sqrt{2(d-1)} \ \log \Bigg[ \sqrt{ \Big( \tfrac{2+\sqrt{2}}{2-\sqrt{2}} \Big)\Big( \tfrac{\gamma-1}{\gamma+1} \Big) \Big(\tfrac{2\gamma-\sqrt{2\gamma^2-1}+1}{2\gamma+\sqrt{2\gamma^2-1}-1} \Big)} 
	\left(\gamma+\sqrt{\gamma^2-\tfrac{1}{2}}\right)^{\sqrt{2}} \Bigg],
\\
&\gamma\equiv \coth\left(\frac{\ka\phi}{\sqrt{2(d-1)}}\right) = \coth \left(- \tilde z / \ell \right).\label{tildephi}
\end{split}
\ee
Although it is not possible to invert this function exactly, it is easy to verify that the asymptotics give the correct CSFI results.
Near the the boundary, $|\tilde z| /  \ell \gg1$ ($\gamma \to 1$), we have 
$\tilde\phi \approx -\phi$, hence $\tilde W(\tilde\phi)\sim e^{\frac{\ka \tilde\phi}{\sqrt{2(d-1)}}}$. This is indeed the critical Liouville behavior. Near the singularity,  $\tilde z \to 0$ ($\gamma \to \infty$), Eq.(\ref{tildephi}) give 
$\ka \phi / \sqrt{2(d-1)} \approx e^{- \frac{\ka \tilde\phi}{2\sqrt{d-1}}}\ll1$, hence
$\tilde W(\tilde\phi)\sim e^{\ka \tilde\phi / \sqrt{d-1}}\gg1$, which is a special Liouville singularity.
The new superpotential (and every bulk quantity) can be written as a function of $\tilde \phi$ only implicitly, through $\gamma$, viz.
\be
\tilde W = \frac{(d-1)}{\ka \ell} \gamma\sqrt{\gamma^2-1} ,	\label{dualWapp}
\ee
but with this we can plot the function $\tilde W(\tilde \phi)$; we do so in Fig.\ref{PlotForApp}.

Here the map between fluctuations is very useful. Although the background is a very complicated implicit function of the field $\tilde \phi$, the tensor fluctuations can easily be found explicitly. The SUSY QM superpotential and the Schr\"odinger potential, as always, are found from (\ref{tilazcsra}),
\be
\begin{split}
\tilde{\cal W}_T (\tilde z) = \tfrac{(d-1)}{2\ell}\coth\left(-\tilde z / \ell \right) \ ,
\quad
\tilde{\cal V}_T (\tilde z) = \frac{(d-1)^2}{4\ell^2} + \frac{(d-1)(d-3)}{4\ell^2 \sinh^2\left(- \tilde z / \ell \right)} .
\end{split}
\ee
The Schr\"odinger equation can be solved as usual, but it is easier to use (\ref{SUSYeinfgpsi})
\begin{align*}
\tilde\psi(z) &=\frac{1}{\sqrt{M^2}} Q \psi(z) \Big|_{z=-\tilde z}
\\
		&=\frac{1}{\sqrt{M^2}} \left[ \partial_z + \frac{(d-1)}{2\ell} \coth\left(z / \ell \right)\right]  \psi(z)\Big|_{z=-\tilde z}.
\end{align*}
Using some properties of the hypergeometric function we can write the result 
\begin{equation*}
\begin{split}
\tilde\psi(\tilde z) &=\tilde C \left[ \sinh (\tfrac{ |\tilde z| }{ \ell}  )\right]^{\frac{d+1}{2}} 
		{}_2F_1\left(\tfrac{d-1}{4} -\tfrac{ig\ell}{2} ,  \tfrac{d-1}{4} + \tfrac{i g\ell }{2}  ;  \tfrac{d}{2}  ;  -\sinh^2 ( \tfrac{|\tilde z|}{ \ell} ) \right),
\\
& g=\sqrt{M^2-\tfrac{(d-1)^2}{4\ell^2}}, \qquad \tilde C=\frac{d}{\sqrt{M^2\ell^2}}C
\end{split}
\end{equation*}
Near the singularity, $\tilde z \to 0$, we have 
$\tilde\psi (\tilde z) \approx \tilde C \left( \frac{1}{\ell} |\tilde z|  \right)^{\frac{d-1}{2}}$,
while at the boundary, $|\tilde z| \to \infty$,
\be
\begin{split}
\tilde\psi(\tilde z) &\approx\tilde C |\tilde \lambda| \cos\left[g\left(|\tilde z|- \ell \log2 \right)+ \tfrac{\tilde\delta_g}{2} \right] ,
\\
|\tilde \lambda| &= \frac{2 \Gamma(ig \ell )\Gamma(\tfrac{d}{2}) e^{- i\frac{\tilde\delta_g}{2}} }{ \Gamma\big(\tfrac{d-1}{4} + \tfrac{ig\ell}{2}\big) \Gamma \big(\tfrac{d+1}{4} + \tfrac{ig\ell}{2} \big)}.
\end{split}
\ee
Comparison with (\ref{phshsit1}) gives the relation between the two phase-shifts
\be
\delta_g = \tilde\delta_g - \arctan \left(\frac{2g\ell}{d-1}\right).
\ee

\bibliographystyle{utphys}

\bibliography{ReferencesSFDforDW_Third_Revision_V3_SCol} 

\end{document}